\documentclass[graybox]{svmult}
\pdfoutput=1

\usepackage[T1]{fontenc}
\usepackage{amsmath,amssymb,amsfonts,physics}
\usepackage{bbm}
\usepackage{txfonts}

\usepackage{helvet}         
\usepackage{courier}        
\usepackage{makeidx}         
\usepackage{graphicx}        
\usepackage{multicol}        
\usepackage[bottom]{footmisc}
\usepackage{hyperref}        
\usepackage{soul}            
\usepackage{comment}
\usepackage[dvipsnames]{xcolor}

\usepackage{tikz}
\usetikzlibrary{decorations.markings}
\usetikzlibrary{decorations.pathmorphing}

\hypersetup{colorlinks=true,urlcolor=black,citecolor=RoyalBlue}

\usepackage[square,numbers,sort&compress]{natbib}
\makeindex             
                       
\usepackage[normalem]{ulem}

\begin{document}
	

\title*{Black Holes in Asymptotically Safe Gravity}

\author{Alessia Platania \thanks{Corresponding author}}

\institute{Alessia Platania \at Perimeter Institute for Theoretical Physics, 31 Caroline Street North, Waterloo, ON N2L 2Y5, Canada,\\
Nordita, KTH Royal Institute of Technology and Stockholm University, Hannes Alfv\'ens v\"ag 12, SE-106 91 Stockholm, Sweden\\
\email{aplatania@perimeterinstitute.ca}}

\maketitle

\vspace{-3.5cm}
\abstract{In this chapter we review the state-of-the-art of black holes in asymptotically safe gravity. After a brief recap of the asymptotic safety program, we shall summarize the features of asymptotic-safety-inspired black-hole models that have been constructed in the past by the so-called renormalization group improvement. Specifically, we will discuss static configurations, both in spherically- and axially-symmetric settings, the role played by the cosmological constant, and the impact of the collapse dynamics in determining black-hole configurations realized in Nature. In particular, we will review how quantum gravity could modify the Buchdahl limit and the corresponding conditions to form ultra-compact objects and Planckian black holes. We will then proceed by describing the most recent developments, particularly those aiming at making model building in asymptotic safety more rigorous and free from ambiguities. These include self-consistent and coordinate-independent versions of the renormalization group improvement, and next steps to fill the gap between model building and renormalization group computations in asymptotic safety. Finally, we will focus on a selection of results that have been obtained from first-principle calculations or arguments, within and beyond asymptotic safety. Concretely, we will review the state-of-the-art in determining black-hole entropy in asymptotic safety from a microstate counting, and progress in deriving the quantum-corrected Newtonian potential. We will discuss how in quantum gravity theories linked to a gravitational path integral singularity resolution could be achieved by a dynamical suppression of singular configurations. Finally, we will show that---independent of the specific ultraviolet completion of gravity---asymptotic modifications to Schwarzschild black holes are strongly constrained by the principle of least action at large distance scales.
}

\section*{Keywords} 
Quantum gravity; renormalization group; asymptotically safe gravity; black holes


\newpage

\section{Introduction} \label{sect:introduction}

Unraveling the quantum nature of black holes is among the most important objectives of research in quantum gravity. 
Despite the impressive achievements of Einstein's general relativity, the singularities and instabilities characterizing classical black holes indicate that a more fundamental description ought to take over the classical framework. How such a fundamental theory of gravity looks like is an outstanding open question.

Over the years, several proposals have been put forth. Insofar as quantum gravity lives at extremely high energies---above the Planck scale---testing and discriminating between theories is challenging, and theoretical consistency has become a fundamental guidance in constraining different theories. Among the variety of consistency constraints, recovering a gravitational effective field theory in the infrared (IR) starting from the deep ultraviolet (UV) is arguably one of the most important requirements, and only a few theories have managed to pass this test so far. Among them, asymptotically safe gravity~\cite{Percacci:2017fkn,Reuter:2019byg}  has emerged as a minimal while promising proposal, conjecturing that quantum gravity be described by a quantum field theory (QFT) whose UV behavior is controlled by an interacting fixed point of the gravitational renormalization group (RG) flow. The fixed point acts as an attractor for a subset of RG trajectories, providing a UV completion for the theory and rendering it renormalizable \`a la Wilson. 

The presence of an asymptotically safe fixed point for quantum gravity, akin to the asymptotically-free one in quantum chromodynamics, is responsible for a distinctive hallmark of asymptotic safety: its anti-screening character~\cite{Nink:2012vd}. On the formal side, gravitational anti-screening is tied to the attractivity properties of the fixed point; on the phenomenological side, this hallmark can be encoded in an effective Newton coupling which vanishes in the regimes where quantum-gravity effects are expected to be important. This intuitive picture has been extensively exploited in the literature to model deviations from general relativity induced by an asymptotically safe UV completion of gravity. The corresponding asymptotic-safety-inspired models are typically obtained by replacing the observed Newton constant with an effective, coordinate dependent one, which smoothly interpolates between the observed value in the IR and its fixed-point scaling in the UV. This procedure is better known as RG improvement and the resulting gravitational models are dubbed RG-improved spacetimes. 

The RG improvement has been a valuable instrument to explore possible implications of asymptotically safe gravity in astrophysics and cosmology, particularly at the dawn of asymptotically safe phenomenology. The method has even inspired an entirely new program, which is by now detached from asymptotically safe gravity, and goes under the name of ``scale-dependent gravity''~\cite{Contreras:2013hua,Koch:2015nva,Contreras:2017eza,Contreras:2019cmf}. Yet, more rigorous derivations and arguments, grounded either on the functional integral or on the effective action, are in order to unravel asymptotic-safety-induced modifications of classical black holes and early-universe cosmology. Such fundamental approaches have been the focus of the asymptotic safety program in the past few years.

In this chapter we review of the state-of-the-art of black holes in asymptotically safe gravity (see also~\cite{Koch:2014cqa,Saueressig:2015xua,Eichhorn:2022bgu}), from the early works on RG-improved black holes to the most recent developments involving computations and considerations based on the gravitational effective action. In our narrative we will mostly be following a chronological order. 

The chapter is organized as follows. In Sect.~\ref{sect:as-basics} we will briefly review the asymptotic safety scenario for quantum gravity, providing all basic ingredients required for the understanding of the subsequent sections on asymptotically safe black holes. The ideas behind the RG improvement, its recipe, and its most pressing issues will be the topic of Sect.~\ref{sect:RG-improv-key-idea}. We will discuss the resulting RG-improved black holes, in all their facets, in Sect.~\ref{sect:rg-improved-black-holes}. Attempts to ameliorating the method and removing its ambiguities will be the focus of Sect.~\ref{sect:refined-rg-imp}, where we will also comment on the physical implications of these improvements and how they compare with past results. Sect.~\ref{sect:bh-from-frg} will be devoted to the subject of black holes from the effective action: we will discuss the most important developments of the past years, which unlocked intriguing aspects of black holes within and beyond asymptotic safety, grounded on first-principle calculations in quantum gravity. We will summarize all these points in our conclusions, Sect.~\ref{sect:conclu}.

\section{Asymptotic safety in a nutshell}\label{sect:as-basics}

Asymptotically safe gravity~\cite{Percacci:2017fkn,Reuter:2019byg} is one of the most conservative approaches to quantum gravity. It relies on the framework of QFT, and conjectures that the high-energy behavior of the gravitational RG trajectory realized by Nature be controlled by a UV fixed point, where all (essential) running couplings approach a finite, non-zero value. This condition is known as ``asymptotic safety'', and can be regarded as a non-perturbative generalization of the well-known concept of asymptotic freedom, whereby couplings vanish in the UV. Asymptotic freedom or safety guarantee that a theory be renormalizable, as well as UV complete with respect to a free or interacting fixed point, respectively. The first case can be related to perturbative renormalizability. The second one corresponds to a generalized notion of renormalizability, often regarded as ``non-perturbative renormalizability''. Power-counting arguments only hold for the former, while the existence of the latter cannot be determined a priori, and must be investigated by appropriate RG techniques, typically beyond perturbation theory. One of their analytical realizations is the framework of the FRG~\cite{Dupuis:2020fhh}, while corresponding lattice approaches are employed within the eucliden and causal dynamical triangulation programs~\cite{Laiho:2016nlp,Loll:2019rdj}.
Thanks to these powerful methods, the asymptotic safety conjecture has been tested within a large number of approximations and against a variety of different starting assumptions (see~\cite{Percacci:2017fkn,Reuter:2019byg} and references therein). Notably, the asymptotic safety approach to quantum gravity does not need new physics, at least in principle, as long as its introduction is not required by compelling experimental evidence.

The FRG combines together two ingredients: on the one hand, the Wilsonian idea of renormalization~\cite{Wilson:1973jj} and, on the other hand, the functional approach introduced in QFT to perform a path integral quantization and to study scattering amplitudes. The result of this combination is a functional approach to renormalization, by which one can 
\begin{itemize}
	\item Handle with any QFT, including those that are (perturbatively or power-counting) non-renormalizable but non-perturbatively renormalizable.
	\item Determine the bare action from first principles, i.e., as an RG fixed point.
	\item Derive the quantum effective action stemming from an RG fixed point and an appropriate number of initial conditions.
\end{itemize}
To this end, one needs to introduce a scale-dependent version of the effective action,~$\Gamma_k$, dubbed  the ``effective average action'' (EAA). Here~$k$ is an artificial RG scale and not a physical momentum. It can be shown that the (Euclidean\footnote{FRG computations are typically performed in Euclidean. First steps towards Lorentzian calculations have been taken in~\cite{Manrique:2011jc,Rechenberger:2012dt,Biemans:2016rvp,Biemans:2017zca,Houthoff:2017oam,Platania:2017djo,Baldazzi:2019kim,Knorr:2018fdu,Eichhorn:2019ybe,Bonanno:2021squ}, while Lorentzian computations based on the spectral FRG have been developed and applied in~\cite{Fehre:2021eob,Braun:2022mgx,Kluth:2022wgh}.}) path integral can be translated into the following functional integro-differential equation for the effective average action,
\begin{equation}\label{eq:flow-eq}
	k\partial_k \Gamma_k = \frac{1}{2} \textrm{STr} \left(\left(\Gamma_k^{(2)}+\mathcal{R}_k \right)^{-1}k\partial_k\mathcal{R}_k\right)\,,
\end{equation}
better known as the Wetterich equation~\cite{Wetterich:1992yh,Morris:1993qb}.
Here $\mathcal{R}_k\propto k^2$ is a RG-scale dependent mass term which is added to the original Lagrangian to regularize the path integral and to integrate quantum fluctuations with momenta $p^2 \gtrsim k^2$. The symbol ``STr'' stands for a supertrace, summing over internal indices and integrating over the volume in coordinate or momentum space. Finally, $\Gamma_{k}^{(2)}$ denotes the second functional derivative of the EAA with respect to all fields appearing in~$\Gamma_{k}$.

The RG fixed points are identified by the conditions $k\partial_k g_i(k) = 0$, where the index~$i$ labels the running couplings in $\Gamma_k$, and $g_i(k)$ denotes their dimensionless counterpart. 
A fixed point can be reached by some RG trajectories either in the IR or in the UV. Accordingly, a given fixed point provides a UV completion for all RG trajectories belonging to its basin of attraction. The dimension of the latter determines the number $N$ of relevant directions, and thus the number of free parameters of the theory. Note that this number depends on the specific fixed point considered.

Once one identifies the set of fixed points, the next question to ask is whether any of the RG trajectories departing from a fixed point in the UV can reach an IR that is compatible with experimental and observational data. This can be verified by integrating the flow equation~\eqref{eq:flow-eq} complemented by a sufficient number of initial conditions dictated by observations. If a solution compatible with these initial conditions exists, and if it reaches a fixed point in the UV, then 
\begin{itemize}
	\item The RG trajectory $\Gamma_k^{sol}$ realized by the particular system analyzed is consistent at all scales and in particular it is UV complete.
	\item The theory associated with the RG trajectory $\Gamma_k^{sol}$ is renormalizable.
	\item Its observables, including scattering amplitudes, can be computed using the standard effective action, which is obtained as the limit $\Gamma_0\equiv \lim_{k\to0} \Gamma_k^{sol}$.
	\item If the theory $\Gamma_k^{sol}$ is UV completed by a fixed point with $N$ relevant directions, all infinitely many couplings in the corresponding effective action $\Gamma_0$ will be written in terms of $N$ free parameters only.
\end{itemize}

The next ingredient to discuss is the practical resolution of the Wetterich equation. Albeit this is exact and provides a clear recipe to compute the functional integral, solving it is involved and one has to resort to approximations. In particular, independent of the specific theory---identified by a set of fields and the symmetries of their interactions---the functional $\Gamma_k$ contains infinitely many interaction terms. In order to make computations doable, one possibility is to project the RG flow onto a managable sub-space  of couplings. The calculation can then be improved in a step-by-step fashion, exploring larger and larger sub-spaces. In practice, this projection is performed by expanding $\Gamma_k$ according to a certain criterion (e.g., a derivative or vertex expansion) and by ``truncating'' the resulting series to a certain order. Obtaining the same results independent of the type of expansion and truncation order is considered as evidence of their stability~\cite{Denz:2016qks}.

In the case of gravity a particularly convenient way to express $\Gamma_k$ is via a curvature expansion of the form~\cite{Knorr:2019atm}
\begin{equation}\label{eq:EAA-quadratic}
	\Gamma_k=\int \dd[4]x \sqrt{-g}\left(\frac{1}{16\pi G_k}\qty(R-2\Lambda_k)+R\, g_{R,k}(\Box)\,R+ C_{\mu\nu\sigma\rho}\,g_{C,k}(\Box)\,C^{\mu\nu\sigma\rho}\right) \,.
\end{equation}
Here $\Box=g_{\mu\nu}D^\mu D^\nu$ is the d' Alembertian operator, and $G_k$, $\Lambda_k$, $g_{R,k}(\Box)$ and $g_{C,k}(\Box)$ are the running Newton, cosmological and quartic couplings, respectively. Only the latter two couplings can depend on the d' Alembertian operator, since in the volume term $\Box$ cannot act on any field, while a term $F_k(\Box)R/G_k$ is equivalent to $R/G_k$ modulo a total derivative. Provided that a well-definite UV completion exists, the limit $k\to0$ defines an effective action resembling closely the structure in Eq.~\eqref{eq:EAA-quadratic}~\cite{Knorr:2019atm}
\begin{equation}\label{eq:eff-action}
	\Gamma_0=\int \dd[4]x \sqrt{-g}\left(\frac{1}{16\pi G_N}\qty(R-2\Lambda)+R\,g_R(\Box)\,R+ C_{\mu\nu\sigma\rho}\,g_C(\Box)\,C^{\mu\nu\sigma\rho}\right) \,,
\end{equation}
where $G_N\equiv G_0$ and $\Lambda\equiv\Lambda_0$ are the observed Newton and cosmological constants. The ``form factors''  $g_{R}(\Box)$ and $g_{C}(\Box)$ encode instead the physical momentum dependence of the quartic couplings~\cite{Christiansen:2014raa, Knorr:2018kog, Bosma:2019aiu, Knorr:2019atm, Knorr:2021niv, Bonanno:2021squ, Fehre:2021eob}, generalizing the relation $\partial^2\sim - p^2$ to curved spacetimes~\cite{Knorr:2019atm}. As the effective action~\eqref{eq:eff-action} is the result of integrating over all quantum fluctuations, quantities computed using it already encode all quantum effects. In particular, quantum black holes and cosmologies ought to be derived as solutions to the dressed field equations
\begin{equation}\label{eq:full-eom}
	\frac{\delta \Gamma_0[g_{\mu\nu}^{sol}]}{\delta g_{\mu\nu}^{sol}}=0\,.
\end{equation}
We now have all ingredients to introduce the so-called ``RG improvement'' and then to step into the main topic of this chapter.

\section{RG improvement: key idea}\label{sect:RG-improv-key-idea}

The RG improvement was introduced in the context of gauge and matter field theories as a short-cut to access some of terms in the effective action and, in some cases, determine leading-order modifications to the solutions to the corresponding effective field equations~\cite{Coleman:1973jx,Migdal:1973si,Matinyan:1976mp,Adler:1982jr,Dittrich:1985yb}. 

The RG improvement procedure consists of the following steps. First, starting from a classical system, e.g., an action or a solution, one replaces its couplings with their running counterparts, $g_i\to g_i(k)$. At the level of the action, this step is akin to promoting an ansatz for the action to an EAA $\Gamma_k$. Second, the running couplings~$g_i(k)$ are replaced with the solutions to the corresponding RG equations, $k\partial_k g_i(k)=\beta_i[g_j(k)]$, complemented by suitable physical initial conditions. Finally,~$k$ is identified with a scale of the system which could act as a physical IR cutoff.

\begin{figure}[t]
		\centering 
		\includegraphics[width=0.7\textwidth]{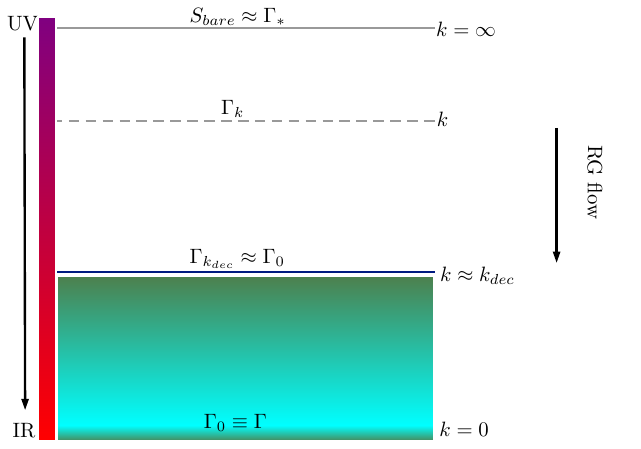}
		\caption{\label{fig:decoupl} Pictorial representation of the decoupling mechanism~\cite{Borissova:2022mgd}. If one or more physical IR scales in the scale-dependent effective action $\Gamma_k$ overcomes the artificial regulator $\mathcal{R}_k$ in Eq.~\eqref{eq:flow-eq}, the flow slows down and eventually freezes out, so that~$\Gamma_{k_{dec}}\approx\Gamma_0$.}
	\end{figure}

The reason why this procedure is supposed to work, at least in simple cases, lies in the so-called \textit{decoupling mechanism}~\cite{Reuter:2003ca}: if in the flow of $\Gamma_k$ there are physical IR scales (e.g., masses, curvature, or interactions terms) that prevail over the unphysical regulator $\mathcal{R}_k$ below a certain threshold scale $k_{dec}$---dubbed the decoupling scale, then the right-hand side of Eq.~\eqref{eq:flow-eq} gets smaller, thus slowing down the RG flow; as a result, the EAA at the decoupling scale approximates the full effective action~$\Gamma_0$. This idea is illustrated in Fig.~\ref{fig:decoupl}.
In particular, the decoupling mechanism can give access to some of the interaction terms in the effective action that were not considered in the initial truncation. This is the case for instance in scalar electrodynamics, where the decoupling condition together with the RG improvement can be used to determine the Coleman-Weinberg effective potential~\cite{Coleman:1973jx,Reuter:2003ca}.

Grounded on the decoupling mechanism, the RG improvement procedure might allow to capture some of the leading-order quantum corrections to classical systems. Yet, its application to gravity is far from straightforward.

First, classical gravitational systems or phenomena are typically characterized by a number of competing physical IR scales, making the ``cutoff identification'' more involved than in the case of standard quantum field theories, where one can na\"ively relate a momentum scale with the inverse of the radial distance. Specifically, despite much effort~\cite{Reuter:2003ca,Babic:2004ev,Domazet:2010bk,Koch:2010nn,Domazet:2012tw,Koch:2014joa}, finding a clear and unique way to identify the cutoff is not straightforward, and has become a source of ambiguities. 

Secondly, the implementation of the RG improvement can in principle be performed at the level of the action, which should be more natural in relation with the decoupling mechanism, or at the level of the field equations or solutions, which is technically less involved and more direct. It is clear however that these procedures can be inequivalent, since an application at the level of the action would yield more terms in the field equations. This is thereby a second source of ambiguity.

Thirdly, in the context of gravity there are two more potential issues: backreaction effects, due to the presence of a dynamical and fluctuating metrics that have to replace the classical, typically singular backgrounds of general relativity, and coordinate independence, which might be lost if the RG improvement is not carefully applied.

In spite of the aforementioned issues, the RG improvement has been a powerful tool to build asymptotic-safety-inspired models and explore possible signatures and consequences of quantum gravity. A selection of these results in black hole physics is the topic of Sect.~\ref{sect:rg-improved-black-holes}. Some recent developments aiming at solving the ambiguities and problems of the RG improvement will be discussed in Sect.~\ref{sect:refined-rg-imp}. Finally, in Sect.~\ref{sect:bh-from-frg} we will report on some findings on quantum black holes stemming from first-principle computations within and beyond asymptotically safe gravity.

\section{RG-improved black holes}\label{sect:rg-improved-black-holes}

In this section we review a selection of asymptotic-safety-inspired black-hole models that have been constructed in the past years via the RG improvement procedure. We will start from the case of spherically-symmetric, asymptotically-flat RG-improved black holes  (Sect.~\ref{sect:bonanno-reuter}). These are the simplest RG-improved models and chronologically the first ones that have been developed.
Next, we will discuss the role of the cosmological constant (Sect.~\ref{sect:cosmolo-const-role}) and the generalization to axially-symmetric systems (Sect.~\ref{sect:rotating-shadows}). We will proceed with a discussion of dynamical RG-improved models, describing the formation of quantum black holes from a gravitational collapse, and we will conclude by detailing how the classical Buchdahl limit could be affected by gravitational antiscreening (Sect.~\ref{sect:gravitational-collapse}).

\subsection{The spherically-symmetric, asymptotically-flat case}\label{sect:bonanno-reuter}

This subsection summarizes the works that pioneered the study of RG-improved black holes~\cite{Bonanno:1998ye,Bonanno:2000ep,Bonanno:2006eu}, and their generalizations in the presence of  extra dimensions~\cite{Burschil:2009va,Falls:2010he}. The focus is the case of static, spherically-symmetric, asymptotically-flat black holes. The RG improvement of the same system and its generalization to the charged case were also subsequently considered in~\cite{Emoto:2005te,Emoto:2006vx,Falls:2012nd,Koch:2013rwa,Koch:2015nva,Gonzalez:2015upa,Zhang:2018xzj,Chen:2022xjk}, whereby the use of different cutoff identifications leads to results in qualitative agreement  with~\cite{Bonanno:1998ye,Bonanno:2000ep,Bonanno:2006eu}.  
Recent developments based on more rigorous versions of the RG improvement and first-principle calculations will be discussed in Sect.~\ref{sect:refined-rg-imp} and Sect.~\ref{sect:bh-from-frg}, respectively.

\paragraph{{\bf Bonanno-Reuter black holes} \cite{Bonanno:1998ye,Bonanno:2000ep}}

Static, spherically-symmetric, asymptotically-flat black holes yield an ideal playground to develop and investigate spacetime models beyond general relativity. In their most commonly studied incarnation the time and radial metric components in Schwarzschild coordinates are assumed to be inversely related, $g_{rr}=g_{tt}^{-1}$, and the line element reads
\begin{equation}\label{eq:line-element}
	ds^{2}=-f(r)\,dt^{2}+f^{-1}(r)\,dr^{2}+r^{2}d\Omega^{2}\,.
\end{equation}
Inasmuch classical spacetimes in general relativity must satisfy the vacuum Einstein field equations, the classical lapse function $f=f_{cl}$ takes the form
\begin{equation}\label{eq:class-class-lapse}
	f_{cl}(r)=1-\frac{2 m G_0}{r}\,,
\end{equation}
where $m$ is the mass of the black hole and $G_0\equiv G_N$ is the observed value of the Newton coupling. Assuming that black holes realized by Nature are such that $g_{rr}=g_{tt}^{-1}$, deviations from general relativity can be embedded in an effective Newton coupling~$G(r)$ modifying the radial dependence of the classical lapse function,
\begin{equation}\label{eq:class-lapse}
	f_{qu}(r)=1-\frac{2 m G(r)}{r}\,.
\end{equation}
In the following we shall analyze the consequences an effective Newton coupling originating from the RG improvement of the classical metric, following the original derivation in~\cite{Bonanno:1998ye,Bonanno:2000ep}.

\begin{figure}[t] 
	\centering\includegraphics[width=0.7\textwidth]{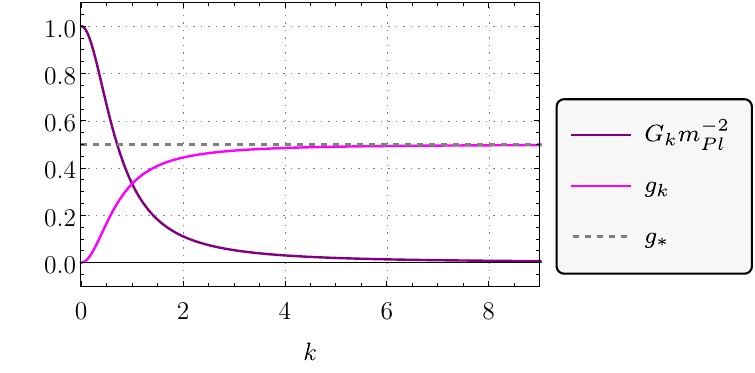}
	\caption{\label{fig:Newt-runnig} Running of the dimesionful (purple line) and dimensionless (magenta line) Newton couplings with the RG scale $k$, according to Eq.~\eqref{eq:running}, with $g_\ast=1/2$ (gray, dashed line). The dimensionful Newton coupling matches its IR value for $k=0$ and vanishes in the UV due to the fixed-point scaling, $G_k\sim g_\ast k^{-2}$. By contrast, its dimensionless counterpart is zero in the IR and approaches its fixed-point value $g_\ast$ as $k\to \infty$.}
\end{figure}

According to the recipe detailed in Sect.~\ref{sect:RG-improv-key-idea}, the RG improvement of a classical metric involves replacing the Newton constant with its running counterpart,
\begin{equation}
	G_0\to G_k\,.
\end{equation}
The functional dependence of $G_k$ on the RG scale $k$ is dictated by the beta function of its dimensionless version, $g_k=G_k k^2$. In an approximation where all other couplings in the action vanish, the running is given by~\cite{Bonanno:1998ye} 
\begin{equation}\label{eq:running}
	G_k=\frac{G_0}{1+g_{\ast}^{-1}G_0 k^2}\,,
\end{equation}
with $g_{\ast}$ being the fixed-point value of the dimensionless Newton coupling $g_k$. The running of the two versions of the Newton coupling is displayed in Fig.~\ref{fig:Newt-runnig}.

Motivated by the case of quantum field theories on flat spacetimes, whereby the Wilsonian IR momentum is inversely related to the radial coordinate, $k\sim 1/r$, in~\cite{Bonanno:1998ye,Bonanno:2000ep} Bonanno and Reuter set the scale via the proper distance between the origin $r=0$ and a generic point along a purely radial geodesic, $k\sim\xi/D(r)$, with $\xi$ a numerical factor (presumably of $\mathcal{O}(1)$) setting the scale of quantum gravity, and the geodesic distance given by
\begin{equation}
	D(r)=\int_0^r d r^{\prime}\left|f_{cl}(r^\prime)\right|^{-1/2}\,.
\end{equation}
It is important to notice that this is only an approximation, as the definition of the proper distance (or any other physical scale) in terms of the classical metric is not self-consistent (cf. Sect.~\ref{sect:iterativeRGimpro}). 

\begin{figure}[t] 
	\centering\includegraphics[width=0.75\textwidth]{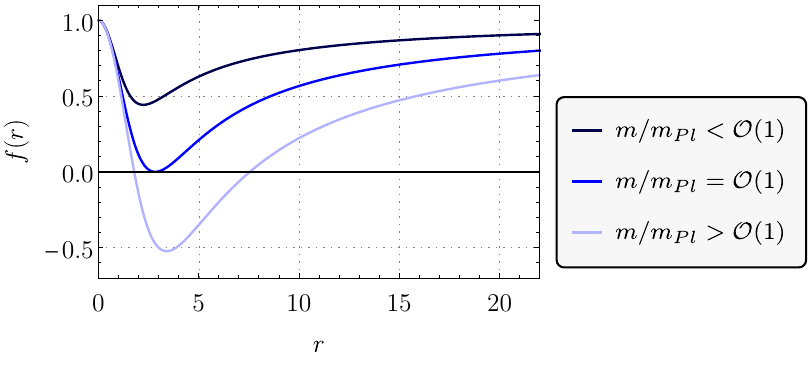}
	\caption{\label{fig:HorizonsBR-BH} RG-improved lapse function~\eqref{eq:BR-lapse} for different values of the black hole mass in Plack units. Above the critical value $m/m_{Pl}\sim 1$ the spacetime has two horizons. If $m$ is decreased, the two horizons get closer, until they merge into a single horizon for $m\sim m_{Pl}$. Below this threshold value, the two horizons disappear, since the roots $r_\pm$ of the Bonanno-Reuter lapse function become complex conjugate.}
\end{figure}

Due to a discontinuity at the classical Schwarzschild radius, $r=r_s=2mG_0$, the analytic form of the proper distance depends on whether $r$ is smaller or bigger than the Schwarzschild radius. Close to the classical singularity it behaves as
\begin{equation}
	D_{r\ll l_{Pl}}(r)\sim\frac{2}{3}\frac{r^{3/2}}{\sqrt{2mG_{0}}}(1+\mathcal{O}(r))\,,
\end{equation}
whereas at large distances it scales as
\begin{equation}
	D_{r\gg l_{Pl}}(r)\sim r +\mathcal{O}(r^0)\,.
\end{equation}
In order to perform analytical calculations, one can exploit an approximate interpolating function,
\begin{equation}\label{eq:cutoffid-propdist}
	D(r)\approx\sqrt{\frac{2r^{3}}{2r+9mG_{0}}}\,,
\end{equation}
that smoothly connects the aforementioned asymptotic behaviors. The RG improvement procedure thereby results in a new black hole whose lapse function reads
\begin{equation}\label{eq:BR-lapse}
	f_{qu}(r)=1-\frac{2mG(r)}{r}=1-\frac{4G_0 m r^2}{2r^3+g_{\ast}^{-1}\xi^2 G_0 (2r+9mG_0)}\,.
\end{equation}
Setting $g_{\ast}^{-1}\xi^{2}={41}/{(10\pi)}$\footnote{The value originally used in~\cite{Bonanno:2000ep} was $g_{\ast}^{-1}\xi^{2}={118}/{(15\pi)}$, and was based on the corrections to the Newtonian potential computed in~\cite{Hamber:1995cq}.}, the asymptotic behavior of the time component of the metric coincides with the leading-order corrections to the Newtonian potential~\cite{Donoghue:1993eb,Bjerrum-Bohr:2002gqz,Khriplovich:2004cx}\footnote{Such an identification is however a stretch, as we will discuss in Sect.~\ref{sect:newpotential}.}. In the opposite regime, close to the would-be singularity at $r=0$, Bonanno-Reuter black holes behave like the most commonly studied regular black holes: they have a de Sitter core with effective cosmological constant
\begin{equation}
	\Lambda_\mathrm{eff}=\frac{4}{3g_{\ast}^{-1}\xi^{2}G_{0}}\,.
\end{equation}
As is typical for regular black holes, Bonanno-Reuter spacetimes can display two, one, or no horizons, depending on the value of the mass in Planck units (cf. Fig.~\ref{fig:HorizonsBR-BH}).

\begin{figure}[t] 
	\centering\includegraphics[width=0.35\textwidth]{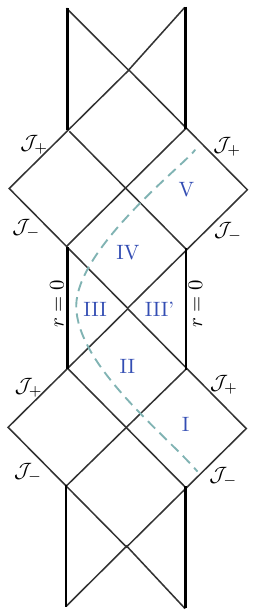}
	\caption{\label{fig:penrose} Penrose diagram of the Bonanno-Reuter black hole~\eqref{eq:BR-lapse}.}
\end{figure}

The causal structure of spacetime is similar to that of a classical Reissner-Nordstr\"om black hole. A test particle departing from the region I ($r<\infty$) will cross the outer horizon $r_+$, thus entering the zone enclosed by the two horizons (region II). Due to the inverse sign of the time component of the metric, particles' geodesics proceed towards the inner horizon $r_-$ and cross it, reaching region III. At this point, however, the sign of the lapse function becomes negative again, and geodesics cross back the inner horizon, in the opposite direction, and land in region IV (see Penrose diagram in Fig.~\ref{fig:penrose}). 

In spite of a similar causal structure, gravitational anti-screening could impact the phenomenon of mass inflation. Classically, due to influx and outflux of gravitational waves during the collapse of a star, the source term for the variation of the mass increases and blows up at the Cauchy horizon, where the spacetime develops a null singularity. The  antiscreening of gravity at high energies can in principle limit the increase rate of the mass function, and could at least weaken the singularity classically formed at the Cauchy horizon~\cite{Bonanno:1998ye}. However, the question of mass inflation and inner horizon (in)stability is still under heated  debate~\cite{Poisson:1989zz,Carballo-Rubio:2018pmi,Bonanno:2020fgp,Carballo-Rubio:2021bpr,Barcelo:2022gii,Carballo-Rubio:2022kad,Bonanno:2022jjp,Carballo-Rubio:2022twq}.

\paragraph{{\bf Evaporation of Bonanno-Reuter black holes} \cite{Bonanno:2006eu}}

The evaporation of Bonanno-Reuter black holes has been studied in detail by the same authors in~\cite{Bonanno:2006eu}, although some features could be guessed by the static limit of the previous section.

\begin{figure}[t] 
	\centering\includegraphics[width=0.65\textwidth]{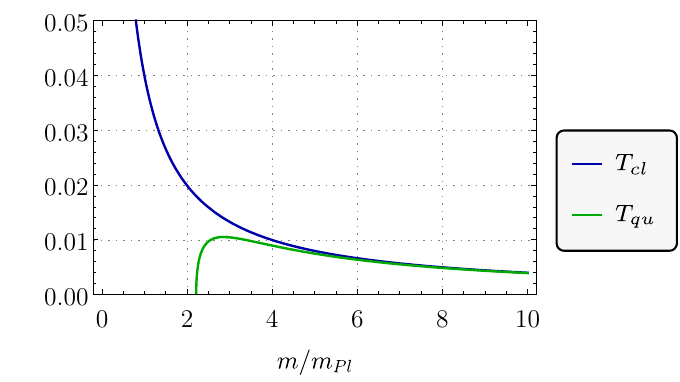}
	\caption{\label{fig:tm} Black hole temperature against the ADM mass $m$ for classical (blu line) and RG-improved (green line) black holes. In the classical case $T\propto 1/m$ and thus the black hole evaporates completely. As for Bonanno-Reuter black holes, the presence of two horizons causes the temperature to vanish at the critical mass $m_{cr}\sim m_{Pl}$. Accordingly, the evaporation process terminates after a finite amount of time and leaves behind a cold Planckian remnant.}
\end{figure}

In general, a key element in determining the endpoint of the evaporation process of a (classical or quantum) black hole is its temperature---defined as the surface gravity at the outer horizon, $r_+$. In the case of Bonanno-Reuter black holes it reads
\begin{equation}
	T=\left.\frac{1}{4\pi}\frac{\partial f_{qu}(r)}{\partial r}\right|_{r=r_{+}}=\frac{1}{8 \pi} \frac{m G_0 r_{+}\left(r_{+}^3-g_{\ast}^{-1} G_0 r_{+}-g_{\ast}^{-1} G_0^2 9 m\right)}{\left(r_{+}^3+g_{\ast}^{-1}\left(r_{+}+9 / 2\, G_0 m\right)\right)^2}\,.
\end{equation}
Starting from the sub-critical configuration with two horizons, the evaporation process makes the temperature increase as $m$ decreases, in analogy with the classical case. However, at variance of classical evaporating black holes, where $T\propto 1/m$ for all values of the black-hole mass, the temperature of Bonanno-Reuter black holes decreases and eventually vanishes as the two horizons collapse into one (see Fig.~\ref{fig:tm}). The evaporation process thus stops, leading to a Planckian black-hole remnant. 

\begin{figure}[t] 
	\centering\includegraphics[width=0.75\textwidth]{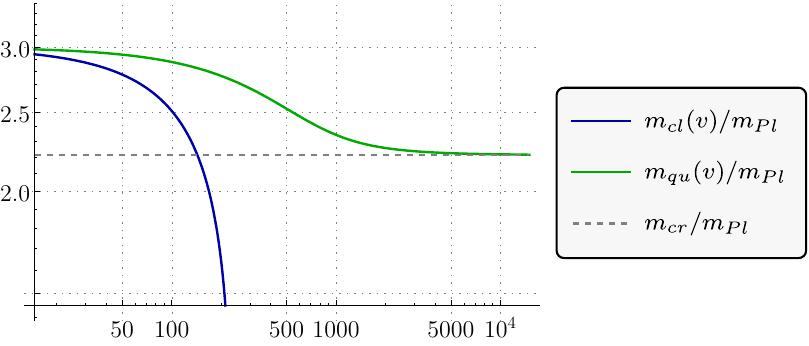}
	\caption{\label{fig:Loglogmv} Dynamics of the black hole ADM mass $m(v)$ in the classical (blue line) and RG-improved (green line) case. The evolution is shown via a log-log plot. For Schwarzschild black holes the mass drops to zero in a finite amount of time, whereas Bonanno-Reuter black holes shrink to a finite size and only reach the final static configuration as $v\to\infty$.}
\end{figure}

The dynamics of the whole process can be studied in more detail by exploiting a generalized Vaidya metric in Eddington-Finkelstein coordinates. Indeed, the evolution of the time-dependent mass $m(v)$ is dictated by the dynamical equation
\begin{equation}
	\dot{m}(v)=-L[m(v)]\,,
\end{equation}
where the black hole luminosity $L$ (energy flux from the outer horizon of the black hole) is given by Stefan-Boltzmann law, $L=\sigma (4\pi r_+^2) T^4$. The result of the numerical integration for the classical and Bonanno-Reuter black holes is shown in Fig.~\ref{fig:Loglogmv}.
The important feature distinguishing Bonanno-Reuter (and, more generally, regular) black holes from their classical counterpart is the timing of the evaporation process: the final remnant configuration is only realized asymptotically, while the evaporation of Schwarzschild black holes occurs in a finite amount of time.  

Thermodynamics of asymptotically safe black holes has been analyzed in great detail in~\cite{Falls:2012nd}. Despite the different starting point, involving an RG improvement at the level of thermodynamical quantities (e.g., the black hole entropy, $S\to S_{k}=\frac{A}{4G_{k}}$) and the use of a different, ``optimized '' cutoff $k=k_{\text{{opt}}}(m,J,Q)\propto1/A\sim1/r_{s}^{2}$ which can in principle depend on the angular momentum $J$ and charge $Q$ of the black hole, the results are in qualitative agreement with those presented above.

Finally, it is worth mentioning that Bonanno-Reuter black holes have been successfully used in the literature to investigate potential implications of quantum gravity. Particularly, quasi-normal modes of Bonanno-Reuter black holes have been studied in~\cite{Rincon:2020iwy}, and it was shown that for Planckian black holes quasi-normal modes display important deviations from the classical case.

After the pioneering work of Bonanno and Reuter~\cite{Bonanno:1998ye,Bonanno:2000ep,Bonanno:2006eu}, several extensions and applications within the spherically-symmetric case have been considered. In the following we briefly summarize modifications induced by the inclusion of extra dimensions.

\paragraph{{\bf Extra dimensions}~\cite{Burschil:2009va,Falls:2010he}}

The derivation of Bonanno-Reuter black holes relies on the physical input that the number of large dimensions is $d=4$. Their generalization to $d\geq 4$ dimensions is relevant in quantum gravity models with large extra dimensions~\cite{Randall:1999vf}; this was the focus of the studies in~\cite{Burschil:2009va,Falls:2010he}. If our universe were $d$-dimensional, the Planck mass $m_{Pl}^2=V_{d-4}m_d^{d-2}$ would be affected by the 
compactified volume of extra dimensions, $V_{d-4}$, and the lapse function of Schwarzschild-like black holes would read
\begin{equation}
	f_{cl}(d,r)=1-\frac{2mG_0}{r^{d-3}}\,.
\end{equation}
The $d$-dimensional Schwarzschild radius, $r_s(d)=\sqrt[d-3]{2mG_0}$, would thus be smaller than in general relativity. 
Interestingly, in the large radii limit the integrand defining the proper distance $D(r)$ is independent of $d$ and thus $D(r)\sim r$ asymptotically for any spacetime dimension. The $d$-dimensional analog of Eq.~\eqref{eq:cutoffid-propdist} is
\begin{equation}
	D(r)\approx\frac{2 r^{\frac{d-1}{2}}}{(d-1)}\left(r_{s}+\left(\frac{d-1}{2}\right)^{-\frac{2}{d-3}} r\right)^{\frac{3-d}{2}}
\end{equation}
and, following the procedure to derive Bonanno-Reuter black holes, one can use this analytic approximation to set the scale, $k\sim 1/D(r)$. This yields a metric with lapse function
\begin{equation}
	f_{qu}(d,r)=1-\frac{2mG_0}{r^{d-3}}\frac{ r^{\frac{(d-1)(d-2)}{2}}}{r^{\frac{(d-1)(d-2)}{2}}+g_\ast^{-1} G_0 \left(\frac{d-1}{2}\right)^{d-2} \left(r_{s}+\left(\frac{d-1}{2}\right)^{-\frac{2}{d-3}} r\right)^{\frac{d^2+6-5d}{2}}} \,.
\end{equation}
The horizon structure and thermodynamical properties of this class of black holes in higher dimensions are analogous to those of Bonanno-Reuter black holes, with two horizons and similar dynamical evaporation.

While the existence of extra dimensions remains unproven, and their theoretical impact of limited use, the presence of a positive cosmological constant is currently strongly supported by observations. Its role in the context of asymptotically safe black holes is the focus of the next subsection. 

\subsection{The role of the cosmological constant}\label{sect:cosmolo-const-role}

Asymptotically-flat RG-improved black holes have been extensively studied in the literature, with the result that gravitational antiscreening yields at least a weakening of the classical singularity. In this subsection we discuss how singularity resolution may be affected by the introduction of a cosmological constant~\cite{Koch:2013owa,Adeifeoba:2018ydh} and by higher-derivative terms in the action~\cite{Cai:2010zh}. 

\paragraph{{\bf Conditions for singularity resolution and the cosmological constant} \cite{Koch:2013owa,Adeifeoba:2018ydh}}

In the case of asymptotically de Sitter spacetimes, the modified mass function reads
\begin{equation}\label{eq:lapse-schw-de-sitter}
	f_{qu}(r)=1-\frac{2m\,G_{k}}{r}-\frac{1}{3}\Lambda_{k}r^{2}\,,
\end{equation}
where $G_k$ and $\Lambda_{k}$ are the dimensionful running Newton and cosmological couplings, and $k=k(r)$. This class of RG-improved black holes was first considered in~\cite{Koch:2013owa}, where the authors argued that the introduction of a cosmological constant $\Lambda_k\neq0$ could be problematic in non-unimodular settings. The argument goes as follows. In an asymptotically safe regime, in $d=4$, the running Newton and cosmological couplings scale as
\begin{equation}\label{eq:tree-level-lg}
	G_k\sim g_\ast k^{-2} \,,\qquad \Lambda_k\sim \lambda_\ast k^2\,.
\end{equation}
The momentum $k$ is then to be related to physical quantities, such as curvature invariants. On dimensional grounds and due to spherical symmetry, the functional relation $k(r)$ is universal in the UV (modulo a numerical factor of $~\mathcal{O}(1)$),
\begin{equation}\label{eq:sc32}
	k\sim D(r)^{-1}\propto K(r)^{1/4} \propto r^{-3/2}\,,
\end{equation}
where $K(r)$ is the Kretschmann scalar. 
Consequently, close to the classical singularity, any scale identification maps the classical Schwarzschild-de Sitter black hole with Newton coupling $G_0$ and cosmological constant $\Lambda_0$ into another Schwarzschild-de Sitter solution with Newton coupling~$\tilde{G}\simeq 3\lambda_*/4 G_0$ and cosmological constant~$\tilde{\Lambda}\simeq 4g_{\ast}/(3G_0)$. 
As a result, unless~$\lambda_\ast\neq0$, the inclusion of a cosmological constant $\Lambda_0$ seems both to reintroduce the singularity and also to restore the thermodynamical properties of classical Schwarzschild-de Sitter black holes \cite{Contreras:2013hua,Koch:2013rwa}\footnote{In a unimodular approach to quantum gravity the cosmological constant emerges as an integration constant instead of a coupling, and thus it does not re-introduce the Schwarzschild singularity~\cite{Torres:2017ygl}.}. 

Since a UV-vanishing cosmological constant would restore the regularity of RG-improved black holes, a key question is whether $\lambda_\ast=0$ is compatible with the spacetime being asymptotically de Sitter, i.e., with an ``emergent'' positive cosmological constant. This question has to do with the conditions for black-hole singularity resolution analyzed in~\cite{Adeifeoba:2018ydh}. 

On dimensional grounds, the leading-order scaling of the IR momentum cutoff~$k(r)$ close to the classical singularity of a spherically-symmetric black hole is
\begin{equation}
	k(r)=\xi\,(mG_{0})^{\gamma-1}r^{-\gamma}\,,
\end{equation}
where $\gamma>0$ by consistency. Although in the spherically-symmetric case dimensional analysis and physical consideration imply $\gamma=3/2$ (cf.~Eq.~\eqref{eq:sc32}) in the following we will keep it unspecified. 

In order to determine the conditions for black-hole singularity resolution, it is convenient to rewrite the lapse function as
\begin{equation}
	f_{qu}(r)=1-2\Phi(r)=1-2\Phi_{G}(r)-2\Phi_{\Lambda}(r)\,,
\end{equation}
i.e., in terms of the ``pseudo Newtonian potential'' $\Phi(r)$ (see also Sect.~\ref{sect:newpotential}) and its components
\begin{equation}
	\Phi_{G}(r)=\frac{mG(r)}{r}\,,\qquad\Phi_{\Lambda}(r)=\frac{\Lambda(r)}{6}r^{2}\,,
\end{equation}
with $G(r)=G[k(r)]$ and $\Lambda(r)=\Lambda[k(r)]$.
The corresponding Ricci and Kretschmann scalars, 
\begin{equation}
		R=\frac{4 \Phi}{r^2}+\frac{8 \Phi^{\prime}}{r}+2 \Phi^{\prime \prime}\,, \qquad K=\frac{16 \Phi^2}{r^4}+\frac{16 \Phi^{\prime 2}}{r^2}+4 \Phi^{\prime \prime 2}\,,
\end{equation}
receive contribution from three quantities: $\Phi/r^{2}$, $\Phi'/r$ and $\Phi''$. The two invariants are thereby regular at $r=0$ if the leading-order scaling of $\Phi(r)$ close to the classical singularity is $\Phi(r)\sim r^{\delta}$
with $\delta\geq2$. This condition can in turn be translated into constraints on $\gamma$ and on the RG flow of the dimensionless Newton coupling $g_k=G_k k^2$ and cosmological constant $\lambda_k=\Lambda_k k^{-2}$~\cite{Adeifeoba:2018ydh}. 

The behavior of $g_k$ and $\lambda_k$ in the proximity of the fixed-point regime~\eqref{eq:tree-level-lg} is obtained by linearizing their beta functions about the fixed point. For a general set of couplings $\{g_i(k)\}$ the linearized beta functions read
\begin{equation}
	k \partial_k g^i(k)=\sum_{j=1}^n \left. \frac{\partial \beta^i}{\partial g^j}\right|_{\mathbf{g}_*}\left(g^j(k)-g_*^j\right)+\mathcal{O}\left(g^j(k)-g_*^j\right)^2\,,
\end{equation}
and, to linear order, the solution takes the form
\begin{equation}
	g^i(k)=g_*^i+\sum_{j=1}^n c_j^i V_j^i\left(\frac{k}{M_{\mathrm{P}}}\right)^{-\theta_j}\,.
\end{equation}
In this expression $c_i^j$ are integration constants selecting the specific RG trajectories, $\theta_i$ are critical exponents, and $V_j^i$ are the components of the corresponding eigendirections. Positive critical exponents thus correspond to IR relevant directions and, vice versa, negative critical exponents identify the irrelevant ones. Similarly to the case of perturbative gauge or matter theories, where one drops out the irrelevant operators in order to select the subset of RG trajectories that are asymptotically free, here we can select the integration constants in front of irrelevant deformations to zero; this choice selects the RG trajectories belonging to the basin of attraction of the fixed point $\{g_*^i\}$. Yet, at variance of perturbative theories, since $g_*^i$ is generally non-zero, selecting the UV critical surface of the fixed point is not equivalent to discarding operators associated with irrelevant directions in the (bare) action.

Restricting ourselves to the Einstein-Hilbert subspace, which is the object of the investigations in~\cite{Adeifeoba:2018ydh}, the scaling of $g_k$ and $\lambda_k$ close to the fixed point reads
\begin{subequations}
	\begin{align}
		&g_k=g_*+g_1\left(\frac{k}{M_{\mathrm{P}}}\right)^{-\theta_1}+g_2\left(\frac{k}{M_{\mathrm{P}}}\right)^{-\theta_2}\,,\\
		&\lambda_k=\lambda_*+\lambda_1\left(\frac{k}{M_{\mathrm{P}}}\right)^{-\theta_1}+\lambda_2\left(\frac{k}{M_{\mathrm{P}}}\right)^{-\theta_2}\,,
	\end{align}
\end{subequations}
and is crucially determined by the critical exponents $\theta_i$.
Using these expressions one can easily show that $\Phi_{G}$ gives a non-singular
contribution to the curvature invariants for $\gamma\geq3/2$~\cite{Adeifeoba:2018ydh}. Moreover, the requirement of singularity resolution yield constraints on the critical exponents $\theta_i$~\cite{Adeifeoba:2018ydh}. 

The first constraint arises from the form of $\Phi_{G}$: the critical exponents are required to be positive.
This is intuitive, since singularity resolution is associated with an effective weakening of the gravitational interaction at high energies. This weakening can come from a gravitational antiscreening only if the theory is asymptotically safe, i.e., if the critical exponents associated with $G$ and $\Lambda$ are positive. 

Additional constraints come from $\Phi_{\Lambda}$: a UV-attractive non-trivial fixed point of the gravitational RG flow implies a UV-scaling for the cosmological constant of the form $\Lambda_{k}\sim\lambda_{\ast}k^{2}$. Although geodesic completeness of an RG-improved Schwarzschild-(A)dS black
hole requires $\lambda_{\ast}$ to be zero~\cite{Koch:2013owa}, if one assumes the spacetime to be asymptotically de Sitter, a positive cosmological
constant ought to re-emerge dynamically. Such a re-emergence is possible in principle, since the corrections to the fixed-point scaling~\eqref{eq:tree-level-lg} become important away from the fixed point and can drive $\Lambda_k$ to a non-zero value. Not all critical exponents would however be compatible with both singularity resolution and the emergence of a positive cosmological constant. Indeed, these physical conditions imply that the critical exponents should  satisfy the inequality $\theta_{i}\geq2$. In turn, this bound is in agreement with many FRG computations, e.g.,~\cite{Biemans:2017zca,Alkofer:2018fxj,Bonanno:2018gck}, and ensures that the cosmological constant could vanish fast enough as $r\to0$ as not to re-introduce a curvature singularity.

\paragraph{{\bf Impact of higher derivatives} \cite{Cai:2010zh}}

Given the general form of the effective action~\eqref{eq:eff-action}, checking the stability of the results obtained within the Einstein-Hilbert truncation against the introduction of higher-derivative terms is of crucial importance. As a first step, Cai and Easson considered an effective action of the form~\eqref{eq:EAA-quadratic} with constant form factors $g_{i,k}(\Box)\equiv g_{i,k}$, and truncated it to quadratic order~\cite{Cai:2010zh}. In such a truncation the class of allowed spacetimes is in principle much richer~\cite{Lu:2015cqa}, but still includes Schwarzschild-like solutions. To determine the impact of higher derivatives on Bonanno-Reuter black holes,~\cite{Cai:2010zh} imposed the starting metric to be of the Schwarzschild-de Sitter form. 
The scale-dependent lapse function is then given by Eq.~\eqref{eq:lapse-schw-de-sitter}, and the RG improvement involves two couplings as before. However, at variance of the analysis in~\cite{Bonanno:1998ye,Bonanno:2000ep,Bonanno:2006eu,Koch:2013owa,Adeifeoba:2018ydh}, where on dimensional grounds $k(r)\sim r^{-3/2} $ (cf. Eq.~\eqref{eq:sc32}), the presence of higher-derivatives introduces new mass scales that can in principle modify Eq.~\eqref{eq:sc32}, or at least render it non-trivial. Specifically, the cutoff function can be determined by the condition that the trace of the generalized Einstein tensor $\tilde{G}_{\mu\nu}$ (which includes contributions from higher derivatives up to quadratic order) be zero, i.e., the modified field equations to be satisfied~\cite{Cai:2010zh}
\begin{equation}\label{eq:gen-cutoffid-R2}
	\mathrm{Tr}\left(\tilde{G}_{\mu\nu}\right)=\frac{k^2}{8\pi g_k}\left(4\lambda_kk^2-R\right)-6 g_R \Box R=0\,.
\end{equation}
This strategy is along the lines of the cutoff identification by Bianchi identities that was formally devised and applied in~\cite{Reuter:2004nv,Reuter:2004nx,Babic:2004ev,Domazet:2012tw,Koch:2014joa}, and that we will discuss in more detail in Sect.~\ref{sect:bianchi}. This procedure is more rigorous than the one employed in~\cite{Bonanno:1998ye,Bonanno:2000ep,Bonanno:2006eu}. On the other hand, this strategy is not applicable in the pure Einstein-Hilbert truncation, in which the above relation becomes trivial. 

Plugging the expression of the curvature invariants and running couplings in Eq.~\eqref{eq:gen-cutoffid-R2}, one finds $k\propto r^{-3/4}$~\cite{Cai:2010zh}, leading to a black hole whose central singularity is weaker than in the classical case albeit not fully resolved. Other than the nature of the singularity, Cai-Esson black holes share all properties of Bonanno-Reuter black holes, from the number of horizons to their thermodynamics. At the same time, the different scaling enforced by the validity of Bianchi identities in the presence of higher-derivatives removes the issues induced by a cosmological constant in the resolution of black hole singularities in the Einstein-Hilbert truncation.

Having discussed static and spherically-symmetric configurations, as well as the role of the cosmological constant, we are now ready to add another key ingredient into the game: rotation.

\subsection{Rotating RG-improved black holes and their shadows}\label{sect:rotating-shadows}

This subsection discusses RG-improved black holes with non-vanishing angular momentum and their shadows~\cite{Reuter:2006rg,Reuter:2010xb,Held:2019xde} (see also~\cite{Eichhorn:2022bgu}). Some variations on the topic, mostly based on different scale identifications, can be found in~\cite{Pawlowski:2018swz,Ishibashi:2021kmf}.

\paragraph{{\bf Reuter-Tuiran rotating black holes}~\cite{Reuter:2006rg,Reuter:2010xb}}

We start by reviewing one (and the first) derivation of the RG-improved metric of a rotating black hole. Spinning black holes are classically described by the Kerr metric. In Boyer-Lindquist coordinates it reads
\begin{equation}\label{eq:metric-kerr}
	\begin{aligned}
		d s^2=&-\frac{\Delta-a^2 \sin^2\theta}{\rho^2} d t^2
		+\frac{\rho^2}{\Delta} d r^2\\
		&+\frac{\left(a^2+r^2\right)^2-a^2 \Delta \sin^2\theta}{\rho^2} \sin^2\theta\, d \phi^2 \\
		&+\rho^2 d \theta^2 -\frac{2\left(a^2+r^2-\Delta\right)}{\rho^2} a \sin^2\theta\, d t d \phi .
	\end{aligned}
\end{equation}
where $a\equiv J/m\in [0,m/m_{Pl}^2]$ is the specific angular momentum and
\begin{equation}
	\begin{aligned}
		\rho^2 =r^2+a^2 \cos ^2 \theta \,, \qquad 
		\Delta =r^2+a^2-2 G_0 m r \,.
	\end{aligned}
\end{equation}
Kerr metrics constitute a 2-parameter family of solutions, the two parameters being the black hole mass $m$ and its angular momentum $J$. A non-zero angular momentum also implies that spherical symmetry turns into axial symmetry with respect to the rotation axis. 

In~\cite{Reuter:2006rg,Reuter:2010xb} the authors followed the Bonanno-Reuter procedure closely, and exploited the proper distance to set the scale. The proper distance from a given spacetime point to the origin along a radial geodesic is now a function of $\theta$,
\begin{equation}
	D(r,\theta)=\int_0^r d\bar{r} \sqrt{ \, \left| \frac{\bar{r}^2+a^2\cos^2\theta}{\bar{r}^2+a^2-2mG_0\bar{r}} \right| \,}\,.
\end{equation}
This function can generally be obtained numerically, while an analytic expression is only possible by restricting $\theta$, e.g., to the equatorial plane $\theta=\pi/2$. As the $\theta$-dependence is rather weak, \cite{Reuter:2006rg,Reuter:2010xb} neglected it to derive the qualitative features of the resulting spacetime. This can be done, for instance, by setting $\theta=\pi/2$. 

For sufficiently large masses (away from the extremality condition), the spacetime resulting from replacing $G_0\to G[d(r)^{-1}]$ in the classical Kerr metric is characterized by two infinite redshift surfaces located at $r_{S_{\pm}}(\theta)$ such that~$g_{tt}=0$, i.e.,
\begin{equation}
	r^2-2G(r)mr+a^2\cos^2\theta=0\,,
\end{equation}
the outer one being the static limit surface. In addition, the Reuter-Tuiran spacetime features two horizons whose radii $r_{\pm}$ are the solutions to the equation $g_{rr}=0$, i.e.,
\begin{equation}
	r^2-2G(r)mr+a^2=0\,.
\end{equation} 
These values are to be found numerically.
Similarly to the case of spherically-symmetric spacetimes, the two horizons coalesce when the black hole mass $m$ is decreased to a threshold value of the order of the Planck mass. In this process, also the two infinite redshift surfaces merge and then disappear.

The Reuter-Tuiran black hole is not a solution to the vacuum Einstein equations, but rather to modified Einstein equations of the form
\begin{equation}
	G_{\mu\nu}=8\pi G_0 T_{\mu\nu}^{\mathrm{eff}}\,,
\end{equation} 
with an effective energy-momentum tensor $T_{\mu\nu}^{\mathrm{eff}}$. As a consequence, the bare black hole mass $m$ and angular momentum $J$ are replaced by corresponding RG-improved quantities, whose value can be computed via the Komar integrals
\begin{subequations}
	\begin{align}
		&M_{qu}^\infty=-(8\pi G_0)\oint \nabla^\alpha t^\beta d S_{\alpha\beta} \,,\\ &J_{qu}^\infty=(16\pi G_0)\oint \nabla^\alpha \phi^\beta d S_{\alpha\beta} \,,
	\end{align}
\end{subequations}
In the expressions above, $t_{\beta}$ and $\phi_\beta$ are the Killing vectors associates with the invariance of the system under time translations and rotations around the spinning axis, respectively. Moreover, the integration is over a two-sphere $S$ at spatial infinity, so that $S_{\alpha\beta}=-2 n_{[\alpha}r_{\beta]}\sqrt{\sigma}d^2\theta$, where $n_\alpha$ and $r_\alpha$ are timelike and spacelike normal vectors to $S$, $\sigma$ is the determinant of the spatial metric $\sigma_{\alpha\beta}$ induced by $g_{\mu\nu}$ on $S$, and $d^2\theta=d\theta_1d\theta_2$ with $\theta_i$ being angular coordinates on $S$. Since the RG improved metric reproduces the classical one asymptotically, the Komar integrals above coincide with their classical counterparts. In contrast, if the two-sphere $S$ is the one enclosed by the outer horizon~$r_+$, the modified Komar integrals~\cite{Reuter:2006rg,Reuter:2010xb}
\begin{subequations}
	\begin{align}
		&M_{qu}=\frac{m G\left(r_{+}\right)}{G_0}\left[1-\arctan \left(\frac{a}{r_{+}}\right) \frac{G^{\prime}\left(r_{+}\right)\left(r_{+}^2+a^2\right)}{a G\left(r_{+}\right)}\right]\,, \\
		&J_{qu}=\frac{J G\left(r_{+}\right)}{G_0}+\frac{M^2 r_{+}^2 G^{\prime}\left(r_{+}\right) G\left(r_{+}\right)}{G_0 a}\left[1-\frac{2 M G\left(r_{+}\right)}{a} \arctan \left(\frac{a}{r_{+}}\right)\right]\,,
	\end{align}
\end{subequations}
only reduce to the bare values in the limit $G(r)\to G_0$. Independent of the specific form of $G(r)$, the effective mass $M_{qu}$ is smaller than its classical counterpart, as expected from the gravitational anti-screening. Yet, Smarr's relation between the mass and angular momentum, 
\begin{equation}
	M_{qu}=2\Omega_{qu} J_{qu}+(4\pi G_0)^{-1}\kappa_{qu} A_{qu}\,,
\end{equation}
with the improved surface gravity, area, and angular frequency given by
\begin{subequations}
\begin{align}
	\kappa_{qu}&={(r_+-2mG_0)(G(r_+)+r_+G^\prime(r_+))}{(r_+^2+a^2)^{-1}}\,, \\
	A_{qu}&=4\pi\,(r_+^2+a^2)\,, \qquad
	\Omega_{qu}={a}\,{(r_+^2+a^2)^{-1}}\,,
\end{align}
\end{subequations}
still holds in the same form as in the classical case.

\paragraph{{\bf Black hole shadows}~\cite{Held:2019xde} }

In~\cite{Held:2019xde} the authors considered a class of RG-improved spinning black holes and studied the size and shape of the corresponding black hole shadow. The cutoff identification is set by $k^4\simeq K$, $K$ being the Kretschmann scalar. For a static spacetime $K=48 G_0^2m^2/r^6$ and the RG improvement maps the classical Schwarzschild spacetime into the Hayward metric~\cite{Hayward:2005gi}. 
In the case of rotating black holes, the Kretschmann scalar also depends on the angular momentum $a$ and the angular coordinate $\theta$,
\begin{equation}
	K(r, \theta, a)=\frac{48 G_0^2 m^2}{\left(r^2+a^2 \cos (\theta)^2\right)^6}\left(r^6-15 r^4 a^2 \cos (\theta)^2+15 r^2 a^4 \cos (\theta)^4-a^6 \cos (\theta)^6\right)\,.
\end{equation}
The authors of~\cite{Held:2019xde} have been the first to employ a $\theta$-dependent cutoff identification to RG improve Kerr black holes. However, since the polynomial
\begin{equation}
	-15 r^4 a^2 \cos (\theta)^2+15 r^2 a^4 \cos (\theta)^4-a^6 \cos (\theta)^6
\end{equation}
is not everywhere positive, it could lead to negative $k^2$ and, consequently, to a complex-valued spacetime metric. To avoid this issue, \cite{Held:2019xde} neglected this part of the Kretschmann scalar and carried out the RG improvement by replacing $G_0\to G_k$ in Eq.~\eqref{eq:metric-kerr}, with $k^2\equiv G_0 m r^3/(r^2+a^2\cos(\theta)^2)^3$ (see~\cite{Eichhorn:2021etc,Eichhorn:2021iwq} for generalizations which exploit horizon-penetrating coordinates and do not neglect the angular dependence). As for the spherically-symmetric case, the event horizon is located inside the classical Schwarzschild radius, resulting in a smaller black hole (and thus in a more compact shadow). Moreover, the higher is the specific angular momentum $a$, the stronger the dependence on $\theta$ is. At this point, one can determine the shadow by looking at how light rays are deviated by a black hole. Concretely, this is done by solving their geodesic equation backward, from a localized distant observer (the camera) to a source.
 
The deviation of the resulting shadow from the one expected from a singular, classical rotating black hole depends strongly on the scale at which quantum gravity effects become important. Enforcing quantum gravity effects at large scales clearly magnifies any deviation. The two major deviations consist of
\begin{itemize}
	\item A shrinking of the shadow, due to an event horizon that is smaller than its classical counterpart. This effect is at work for both the spherically- and the axially-symmetric cases.
	\item In the case of spinning spacetimes, the appearance of a characteristic
	``dent''~\cite{Li:2013jra,Bambi:2015kza,Abdujabbarov:2016hnw,Schee:2017hof,Tsukamoto:2017fxq, Held:2019xde,Eichhorn:2021etc,Eichhorn:2021iwq}\footnote{The dent introduced in~\cite{Li:2013jra,Bambi:2015kza,Abdujabbarov:2016hnw,Schee:2017hof,Tsukamoto:2017fxq} and the one discussed in~\cite{Held:2019xde,Eichhorn:2021etc,Eichhorn:2021iwq} are however structurally different, since in the case of~\cite{Held:2019xde,Eichhorn:2021etc,Eichhorn:2021iwq} the scale identification introduces an angular dependence in all metric components that breaks a mathematical property known as ``circularity''. The non-circularity yields a dent whose boundary has a concave piece, as opposed to the one in~\cite{Li:2013jra,Bambi:2015kza,Abdujabbarov:2016hnw,Schee:2017hof,Tsukamoto:2017fxq} which is fully convex.}. The latter is more evident on the equatorial plane $\theta=\pi/2$ of the rotating black hole and its size depends on the scale of quantum gravity. 
\end{itemize}
If quantum gravity sets in at Planckian scales, as is expected on dimensional grounds, these deviations will be practically unobservable (see also~\cite{Eichhorn:2022bgu,Vagnozzi:2022moj}).
Importantly, under certain assumptions~\cite{Eichhorn:2021etc}, these effects seem to be a general feature of black holes beyond general relativity~\cite{Li:2013jra,Bambi:2015kza,Abdujabbarov:2016hnw,Schee:2017hof,Tsukamoto:2017fxq,Held:2019xde,Eichhorn:2021etc,Eichhorn:2021iwq}.

\subsection{Gravitational collapse and improved Buchdahl limit}\label{sect:gravitational-collapse}

So far we have focused on static black hole configurations and their evaporation. A key ingredient discriminating between physical and unphysical configurations is their dynamics, and specifically their formation via a physical process. Among the possible formation processes, gravitational collapse is of utmost importance in the context of astrophysical black holes. Yet, there is no consensus on how to precisely model the collapse---not even at a classical level---and several models have been developed. In the following we shall summarize some of them, together with their quantum-corrected versions and the corresponding physical implications~\cite{Casadio:2010fw,Fayos:2011zza,Torres:2014gta,Torres:2014pea,Torres:2015aga,Bonanno:2016dyv,Bonanno:2017kta,Bonanno:2017zen,Bonanno:2019ilz}. In particular, the focus of the discussions will be \emph{(i)} the potential avoidance of singularities when dynamics is accounted for, \emph{(ii)} the impact of quantum corrections on Penrose's cosmic censorship conjecture, and \emph{(iii)} how the Buchdahl's limit is modified by gravitational antiscreening. 
As for point \emph{(i)}, two classes of RG-improved models have been considered in the literature:
\begin{itemize}
	\item Those where one fixes the radial dependence of the Newton coupling $G(r)$ from the \emph{RG improvement of the classical static system}
	(e.g., to be of the Bonanno-Reuter type) and subsequently studies the dynamics of the mass  $m(v)$ induced by the different radial dependence. We shall refer to these models as \emph{partially dynamical RG improvements}. Examples are those in~\cite{Fayos:2011zza,Torres:2014gta,Torres:2014pea,Torres:2015aga} and we will discuss them first.
	\item Those where the dynamics is encoded in an effective Newton coupling $G(r,v)$ that depends on both the radial and the time coordinates, and whose specific analytical form is derived from the \emph{RG improvement of the classical dynamical system}. We will call models belonging to this class \emph{fully dynamical RG improvements}. Examples are those in~\cite{Casadio:2010fw,Bonanno:2016dyv,Bonanno:2017kta,Bonanno:2017zen,Bonanno:2019ilz} and we will discuss them next.
\end{itemize}
The discussions \emph{(ii)} and \emph{(iii)} will only involve the second class of models~\cite{Bonanno:2016dyv,Bonanno:2017kta,Bonanno:2017zen,Bonanno:2019ilz}.

\paragraph{\textbf{Singularity resolution and partially-corrected collapse dynamics}~\cite{Fayos:2011zza,Torres:2014gta,Torres:2014pea,Torres:2015aga}} 

The models in~\cite{Fayos:2011zza,Torres:2014gta,Torres:2014pea,Torres:2015aga}, within different classical collapse models and approximations, describe the collapsing spacetime with a homogeneous interior surrounded by a Bonanno-Reuter exterior, which is inserted \emph{ad hoc}. 

In the literature, models of radiating collapsing stars haven been considered, starting from the pioneering work of Oppenheimer and Snyder~\cite{Oppenheimer:1939ue}, to the more realistic Vaidya models~\cite{vaidya1951gravitational}, accounting for the outgoing incoherent radiation, and eventually radiating away all the star mass~\cite{Bondi:1964zza}. These models were soon abandoned due to the observation that the radiation emitted undergoes large backreaction effects due to spacetime curvature, and it is infinitely blueshifted in the limit where the mass gets small~\cite{WAUGH1986154}. In this regime the Vaidya solution is no longer a good approximation, as the large amount of backscattered ingoing radiation ought to be accounted for, and is instead neglected by the outgoing Vaidya model. One may however ask the question of whether a modified exterior model reducing the growth of the curvature as $r\to0$ can  avoid this conclusion.

Inspired by the work of Bonanno and Reuter that we described in Sect.~\ref{sect:bonanno-reuter},~\cite{Fayos:2011zza} considered an improved outgoing Vaidya solution,
\begin{equation}
	\mathrm{d} s^2=-\left(1-\frac{2 m(u) G(r)}{r}\right) \mathrm{d} u^2-2 \mathrm{~d} u \mathrm{~d} r+r^2 \mathrm{~d} \Omega^2\,,
\end{equation}
with $G(r)$ being the Bonanno-Reuter effective Newton coupling~\eqref{eq:BR-lapse}, to describe the exterior region of a collapsing radiating star. Thanks to the regularity of the Bonanno-Reuter lapse function and of its derivatives, the curvature does not grow to infinite as $r\to0$, and the problem of the unbound backscattered radiation may be avoided. To show this, it suffices to integrate the radial null geodesics of a backscattered test field,
\begin{equation}\label{eq:rel-bsr}
	\begin{aligned}
		&\frac{\mathrm{d}^2 u}{\mathrm{~d} \lambda^2}+\frac{r m G^{\prime}-m G }{r^2}\left(\frac{\mathrm{d} u}{\mathrm{~d} \lambda}\right)^2=0 \,, \qquad \frac{\mathrm{d}^2 r}{\mathrm{~d} \lambda^2}+\frac{m_{,u}G }{r}\left(\frac{\mathrm{d} u}{\mathrm{~d} \lambda}\right)^2=0\,.
	\end{aligned}
\end{equation}
Classically, the problem of unboundness of backscattered radiation comes from the behavior of $\partial_\lambda u$, which diverges in the limit $r\to\infty$. A first integration of Eq.~\eqref{eq:rel-bsr} yields the general expression
\begin{equation}
	\frac{\mathrm{d} u}{\mathrm{~d} \lambda}=A \exp \left(\int \frac{G m-r m G^{\prime}}{r^2} \mathrm{~d} u\right)\,,
\end{equation}
with $A$ being an integration constant. Inserting the specific expression of the Bonanno-Reuter effective Newton coupling one can easily see that $\lim_{r\to0}\partial_\lambda u=\text{const}$~\cite{Fayos:2011zza}. 

Within these partially-dynamical RG-improved models, the absence of shell-focusing singularities is inherited by the Bonanno-Reuter scaling of $G(r)$. This has been checked both within the collapse model we just discussed, first analyzed in~\cite{Fayos:2011zza}, and via the more known Lemaitre-Tolman-Bondi (LTB)  model~\cite{Torres:2014gta}. The latter is a class of spherically-symmetric solutions consisting of non-interacting particles (named ``dust''), where the gravitational collapse and the formation of a singularity is classically unavoidable. To discuss singularity resolution in this case,~\cite{Torres:2014gta} modeled the star as an homogeneous interior---a spherically-symmetric collapsing objects, parameterized as a sphere of non-interacting dust particles---surrounded by an RG-improved exterior of the Bonanno-Reuter type. Due to the matching conditions at the star surface, key properties of the Bonanno-Reuter scaling function are inherited by the interior metric. Specifically, the class of improved dust interiors in geodesic coordinates is
\begin{equation}
	d s_{int}^2=-d \tau^2+\frac{R^{\prime}(\tau, r)^2}{1+2 E(r)} d r^2+R(\tau, r)^2 d \Omega^2\,,
\end{equation}
where $E(r)$ is an arbitrary function stemming from a partial integration of one of the matching conditions. It represents the total energy per unit mass of the dust particles within a shell of radius $r$. Moreover, the same matching condition defines the function $R$ as the solution
\begin{equation}
	\frac{\dot{R}^2}{2}=\frac{G(r) M(r)}{R}+E(r)\,,
\end{equation}
where $G(r)$ is the Bonanno-Reuter effective Newton coupling and $M(r)$ is the total mass within a sphere of radius $r$. One can at this point demonstrate that if $R^\prime\neq0$, the energy-momentum tensor of the improved solutions is bounded. As a consequence, the geodesics of dust particles undergo a bounce
in the interior region. In addition, the scalar invariants are finite because of the Bonanno-Reuter scaling. The typical shell-focusing singularities of classical LTB models are thus not formed~\cite{Torres:2014gta}.

The result is stable under the inclusion of backreaction effects due to Hawking radiation. This has been checked both using an ingoing Eddington-Finkelstein model which parametrizes the ingoing negative energy
flux of Hawking radiation~\cite{Torres:2014pea}, and via a standard LTB model~\cite{Torres:2015aga}. In the latter case though, in order 
to avoid the shell focusing due to the propagation of ingoing shells of collapsing matter and bouncing shells of backscattered matter, it is key that the collapse occurs fast enough~\cite{Torres:2015aga}. This allows the final object to be void of singularities.

\paragraph{\textbf{Conditions on singularity avoidance from the collapse of dust shells}~\cite{Casadio:2010fw} }

While singularities appear to be avoided when improving classical static solutions, it is not obvious that dynamical black holes formed from a gravitational collapse will be singularity-free. To determine the impact of the collapse dynamics on singularity resolution,~\cite{Casadio:2010fw} studied quantum corrections to dynamical black hole solutions within one of the simplest collapse models---the
Tolman-Lemaitre-Oppenheimer-Snyder model~\cite{Oppenheimer:1939ue}---via a fully dynamical RG improvement. The latter aims at describing the collapse of a thin shell of dust with areal radius $R(r,\tau)$, with $\tau$ being the proper time of comoving observers along the paths of constant $r$. The exterior and interior geometries are described by the line elements
\begin{subequations}
\begin{align}
	&\mathrm{d} s^2_{ext}=-\left(1-\frac{2 m G_0}{R}\right) \mathrm{d} t^2+\left(1-\frac{2 m G_0}{R}\right)^{-1} \mathrm{~d} R^2+R^2 \mathrm{~d} \Omega^2\,,\\
	&\mathrm{d} s^2_{int}=-\mathrm{d} \tau^2+\left(R^{\prime}\right)^2 \mathrm{~d} r^2+R^2 \mathrm{~d} \Omega^2\,.
\end{align}
\end{subequations}
In particular, the dynamics of $R$ is governed by the classical field equations
\begin{equation}
	\frac{\left(\dot{R}^2 R\right)^{\prime}}{R^2 R^{\prime}}=8 \pi G_0 \,\rho\,,
\end{equation}
where $\rho(r,\tau)$ is the dust energy density, which evolves according to the Bianchi identities
\begin{equation}
	\dot{\rho}+\frac{\partial_\tau\left(R^3\right)^{\prime}}{\left(R^3\right)^{\prime}} \rho=0\,.
\end{equation}
In these expressions and in the following a prime denotes derivation with respect to the radial coordinate $r$, whereas a dot stands for the proper time derivative. Imposing the junction conditions at the boundary of the dust ball, at $r=r_s$, yields the constraint $R(\tau)=a(\tau)r_s$. Correspondingly, the matter energy density becomes $\rho=3/(8\pi G_0)(\dot{a}/a)^2$ and the areal radius satisfies the equation
\begin{equation}
	\dot{R}^2=\frac{2}{R}\frac{4 \pi G_0}{3} \int_0^r \rho(\tau, x)\left[R^3(\tau, x)\right]^{\prime} d x\,.
\end{equation}
If one turns quantum gravity effects on, the effective dynamics might be described by equations that are structurally similar, but with $G_0$ replaced by a coordinate-dependent function. In particular, the dominant energy scale of the system in this case is the matter energy density $\rho$, and thus the classical field equation is replaced by an effective one:
\begin{equation}
	\left(\frac{\dot{a}}{a}\right)^2=\frac{8 \pi \rho}{3}  G(\rho)\,.
\end{equation}
The classical Bianchi identities remain instead unmodified. Before proceeding, we note that the procedure exploited above and drawn from~\cite{Casadio:2010fw} is an RG improvement at the level of the field equation. As it will be clarified later on, in Sect.~\ref{sect:refined-rg-imp}, this is less fundamental than an RG improvement at the level of the action, as it disregard additional terms stemming from higher-derivative quantum corrections in the effective action; nonetheless, a scale identification involving the matter energy density $\rho$ is appropriate and consistent with Bianchi identities.

Combining the field equations for $a$ and $\rho$ yields the integral
\begin{equation}
	\int_{\rho\left(\tau_0=0\right)}^{\rho(\tau)} \frac{\mathrm{d} \rho}{\sqrt{24 \pi \rho^3 G(\rho)}}=\tau\,.
\end{equation}
If the integral is convergent as $\rho\to\infty$, then there exist geodesics attaining the singularity at a finite proper time $\tau_\ast$. Singularity avoidance thus requires the divergence of the integral, and this can only occur if the fall off of $G(\rho)$ is sufficiently rapid. Concretely, close to the classical singularity $G_k\sim k^{-\alpha}$, and one can assume $k^2\sim G_k\, \rho$~\cite{Weinberg:2009wa,Casadio:2010fw}, implying that
\begin{subequations}
	\begin{align}
		&\rho\left(\tau \rightarrow \tau_*\right) \sim\left(\tau_*-\tau\right)^{-(2+\alpha)}\,, \\
		&a\left(\tau \rightarrow \tau_*\right) \sim\left(\tau_*-\tau\right)^{(2+\alpha) / 3}\,, \\
		&k\left(\tau \rightarrow \tau_*\right) \sim\left(\tau_*-\tau\right)^{-1}\,, \\
		&G\left(\tau \rightarrow \tau_*\right) \sim\left(\tau_*-\tau\right)^\alpha\,.
	\end{align}
\end{subequations}
If quantum gravity remains perturbative across all scales, i.e. $k^2 G_k\ll1$, then $\alpha\geq 2$ and the classical singularity is not resolved by the  effective modifications above.
Therefore, as a general result, the collapse dynamics makes singularity resolution less straightforward than in the classical, static case, and requires stronger deviations from the classical dynamics, which go beyond the regime where perturbation theory holds. This general expectation is also met by the concrete model devised in~\cite{Bonanno:2016dyv,Bonanno:2017kta,Bonanno:2017zen}, that we review in the following in relation to the cosmic censorship conjecture.

\paragraph{\textbf{Gravitational collapse and cosmic censorship in asymptotic safety}~\cite{Bonanno:2016dyv,Bonanno:2017kta,Bonanno:2017zen}}

After the formulation of the singularity theorems~\cite{1970HP}, Penrose conjectured that curvature singularities ought to always be hidden behind an event horizon~\cite{1969CC}---a ``cosmic censor'' preventing far-away physicists from seeing their theories breaking down and physics loosing its predictive power. 

In order to test Penrose's conjecture,~\cite{Bonanno:2016dyv,Bonanno:2017kta,Bonanno:2017zen}
studied the dynamical process of black hole formation and discussed the singularity structure as well as the dynamical evolution of the event horizon within
a quantum-gravity-corrected Vaidya-Kuroda-Papapetrou
(VKP) model~\cite{Kuroda:1984vkp,Papapetrou:1985vkp}. 

The classical VKP model describes the gravitational collapse through an ingoing Vaidya metric,
\begin{equation}
	 d s^2=-f(r,v) d v^2+2 d v d r+r^2 d \Omega^2\,,
\end{equation}
with a lapse function that depends on the advanced time $v$ via the dynamical mass~$m(v)$,
\begin{equation}
	f_{cl}(r,v)=1-\frac{2m(v)G_0}{r}\,.
\end{equation}
In this model the spacetime is initially (for $v\leq0$) a flat Minkowski background; subsequently, radiation from a nearby massive star is focused towards~$r=0$. This is modeled by a set of ingoing radial null geodesics, that are focused to~$r=0$, and raise the black hole mass from zero at $v=0$ to $m(v)$. Such a shell-focusing classically cause the formation of a curvature singularity at~$r=0$. The  collapse ends when all radiation from the star is radiated away, and $m(v)$ reaches a final constant value, $m(v)=m$. The resulting spacetime is thus a Schwarzschild black hole with mass $m$. Within the VKP model the growth of the mass function is fixed to be linear, $m(v)=\lambda v$, for $v\in[0,\bar{v}]$, with $\bar{v}$ denoting the advance time at the end of the collapse. 

The procedure of RG improvement yields in this case  a generalized Vaidya spacetime~\cite{Wang:1998qx},
\begin{equation}
	f(r,v)=1-\frac{2M(r,v)}{r}\,,
\end{equation}
with generalized mass function $M(r,v)=m(v)G(r,v)$. Generalized Vaidya spacetimes~\cite{Wang:1998qx} are a class of solutions to Einstein-like field equation with an effective energy-momentum tensor having both a null and a non-null component. It reads
\begin{equation}
	T_{\mu\nu} = \mu \,l_\mu l_\nu + \qty(\rho + p)\qty(l_\mu n_\nu+l_\nu n_\mu) + p g_{\mu\nu}\,,
\end{equation} 
where $l_\mu$ and $n_\mu$ are null vectors satisfying the condition $l_\mu n^\mu = -1$, $\mu$ is the radiation energy density provoking the mass variation of the black hole,
\begin{equation}
	\begin{gathered}
		\mu(r, v)=\frac{1}{4 \pi G_0 r^2} \frac{\partial M(r, v)}{\partial v} \,,
	\end{gathered}
\end{equation}
whereas
\begin{equation}\label{eq:rhop-gen-vaidya}
	\begin{gathered}
		\rho(r, v)=\frac{1}{4 \pi G_0 r^2} \frac{\partial M(r, v)}{\partial r},\quad p(r, v)=-\frac{1}{8 \pi G_0 r} \frac{\partial^2 M(r, v)}{\partial r^2} \,,
	\end{gathered}
\end{equation}
are the energy density and pressure associated with the non-null fluid sourcing the effective modification to Einstein equation due to the replacement $G_0\to G(r,v)$. 

In an RG-improved model $G(r,v)\equiv G(k[r,v])$, with the functional form of $G$ given by Eq.~\eqref{eq:running}. The cutoff identification considered in~\cite{Bonanno:2016dyv,Bonanno:2017kta,Bonanno:2017zen} is $k^{4}\sim\rho$, $\rho$ being the energy density of the ingoing radiation. Indeed, since 
the black hole interior is homogeneous and formed by radiation, one can exploit the relation $\mu_{cl} \sim T^4\sim k^4$~\cite{Manrique:2011jc}, where $\mu_{cl}$ is the energy density in the classical ($G(r,v)=G_0$) Vaidya model, 
\begin{equation}
	\mu_{cl}(r)=\frac{\dot{m}(v)}{4 \pi r^2}=\frac{\lambda}{4 \pi r^2}\,.
\end{equation}
The resulting RG-improved lapse function reads~\cite{Bonanno:2016dyv}
\begin{equation}\label{eq:fqudyn}
	f_{qu}(r,v)=1-\frac{2 m(v) G_0}{r+\alpha \sqrt{\lambda}}\,,
\end{equation}
with $\alpha\equiv g_{\ast}^{-1}G_0/\sqrt{4\pi}$. To leading order, the Ricci and Kretschmann scalars scale as $\sim 1/r^2$ and $\sim 1/r^4$, respectively, indicating a weakening of the classical singularity. Specifically, the singularity at $r=0$ turns out to be gravitationally weak (or, integrable, according to Tipler classification \cite{Tipler:1977zzb}): curvature invariants diverge as $r\to0$ but the geodesic equation can be integrated, so that the spacetime is geodesically complete.
The case of black holes with integrable singularities, as the one derived in~\cite{Bonanno:2016dyv} and further explored in \cite{Bonanno:2017kta,Bonanno:2017zen}, is particularly important, as it could provide a consistent alternative to regular black holes whose inner horizon is unstable~\cite{Poisson:1989zz,Carballo-Rubio:2018pmi,Bonanno:2020fgp,Carballo-Rubio:2021bpr,Barcelo:2022gii,Carballo-Rubio:2022kad,Bonanno:2022jjp,Carballo-Rubio:2022twq}.

Due to the additional term in the denominator of the lapse function, the apparent horizon (AH) gets shifted with respect to the classical case,
\begin{equation}
	r_{AH}=2m(v)G_0-\alpha\sqrt{\lambda}\,.
\end{equation}
Although in the case of dynamical spacetimes the AH does not coincide with the event horizon (EH), the above expression indicates that in the quantum-corrected spacetime horizons tend to form later than in the classical version. Specifically, the minimum irradiation period required to form an horizon is ${v}_{min}=(2g_\ast \sqrt{4\pi\lambda})^{-1}$. 

\begin{figure}[t!]
	\centering
	\resizebox{0.5\textwidth}{!}{\begin{tikzpicture}
\tikzset{->-/.style={decoration={markings,mark=at position {#1} with {\arrow{>}}}, postaction={decorate}}}
\tikzset{-<-/.style={decoration={markings,mark= at position {#1} with {\arrow{<}}},
postaction={decorate}}}
\draw[decoration={zigzag, amplitude=0.6mm, segment length = 2mm}, decorate,thick] (-3-1-1.5+0.5,4-4+1) -- (-3+3,4-3) node[pos=.5,sloped,above] {$r=0$};
\draw [decoration={zigzag, amplitude=0.6mm, segment length = 2mm}, decorate,thick] (-3-1-1.5+0.5,4-4+1) -- (3.5-1.5-2.5-1.5-1.5,4-4-2.5+1.5+0.5)  node[pos=.52,sloped,below] {$r=0$};
\draw [decoration={zigzag, amplitude=0.6mm, segment length = 2mm}, decorate,thick](3.5-1.5-2.5-1.5-1.5,4-4-2.5+2) -- (3.5-1.5-2.5-1.5-1.5,4-4-2.5-2);
\draw [thick] (3.5-1.5-5.5,4-4-2.5-2) -- (3.5-1.5-5.5,-2.5-4+1.5-1);
\draw (-3+3,4-3) -- (4-2.25,0.5-1.25) node[pos=.35,right] {$\;\mathcal{J}^+$};
\draw (4-2.25,0.5-1.25) -- (3.5-1.5-5.5,-2.5-4+1.5-1) node[pos=.6,sloped,below] {$v \; \longrightarrow$} node[pos=.17,right] {$\,\mathcal{J}^-$};
\draw [dashed] (3.5-1.5-5.5,4-4-2.5-2) -- (3.5-1.5-5.5+0.75,4-4-2.5-2-0.75) node[pos=.5,sloped,above] {$v=0$};
\draw [dashed] (-3.5,-0.5) -- (-3.5+2.75,-0.5-2.75) node[pos=.87,sloped,above] {$v=\tilde{v}$};
\draw [dashed] (-1.5-1.5,-0.5+1.5) -- (0.25,-2.25) node[pos=.885,sloped,above] {$v=\bar{v}$};
\draw [thick] (3.5-1.5-2.5-3,4-4-2.5-3+0.5+0.5) -- (3.5+0.5-2.5-0.5,-2.5-0.5+2.5+0.5) node[pos=.36,sloped,below] {CH};
\draw [thick] (3.5-1.5-2.5-1.5-1.5,4-4-2.5+1.5-1.5) -- (3.5+0.5-2.5-1.5,-2.5-0.5+2.5+1.5) node[pos=.2,sloped,below] {EH};
\draw [thick] (-3.5,-0.5) to [out=270+55,in=270-50] (-1.5,-0.5);
\node at (-2.55,-0.65) {AH};
\end{tikzpicture}}
	\caption{Penrose diagram of the dynamical spacetime with improved lapse function~\eqref{eq:fqudyn} for $\lambda\leq\lambda_c$. \label{fig:figcon}}
\end{figure}
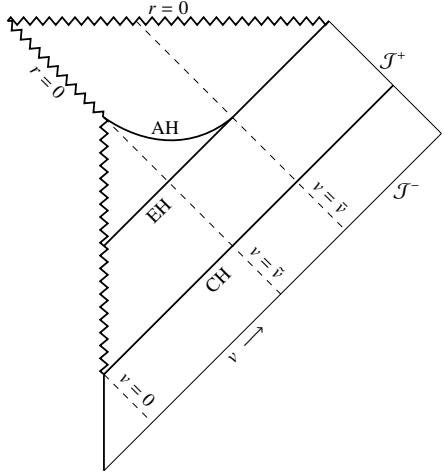

Similarly to the classical collapse~\cite{Kuroda:1984vkp,Papapetrou:1985vkp}, there exists a critical irradiation rate~$\lambda_c$ below which the singularity is formed before the EH. In this case the singularity is globally naked and the first null ray departing from the singularity and reaching future null infinity coincides with a Cauchy horizon. The overall causal structure of the RG-improved spacetime for $\lambda\leq \lambda_c$ is drawn in Fig.~\ref{fig:figcon}. 

A striking feature of the RG-improved model is that the critical infusion rate~$\lambda_c$ is higher than in the classical case. Therefore, the formation of naked singularities and the corresponding violation of Penrose's cosmic censorship conjecture seem to be favored by gravitational antiscreening. This can be understood intuitively, in so far as a weakening of gravity towards high energies can delay the formation of  horizons. At the same time, the replacement of the classical singularity with an integrable one might render the presence of cosmic censor unnecessary.

\paragraph{\textbf{Stellar mass-radius relation and improved Buchdahl limit}~\cite{Bonanno:2019ilz}}

A broader, if perhaps less detailed way to investigate possible endpoints of the gravitational collapse of massive stars is to analyze the equilibrium configuration of massive astrophysical objects. This is classically determined by the so-called Tolman–Oppenheimer–Volkoff (TOV) stellar equilibrium equation~\cite{Tolman:1939jz,Oppenheimer:1939ne}. The derivation of the latter goes as follow. One models a star as a self-gravitating perfect fluid with proper energy density $\epsilon(r)$ and pressure determined by an equation of state $p=p(\epsilon)$. The star energy-momentum thus reads
\begin{equation}
	T_{\mu \nu}=(\epsilon+p(\epsilon)) c^{-2} u_\mu u_\nu-p(\epsilon) g_{\mu \nu}\,,
\end{equation}
and enters the right-hand side of the Einstein equations. Inserting in the resulting modified equations the expression for the most general static and spherically-symmetric metric and combining it with the Bianchi identities gives the TOV equation~\cite{Tolman:1939jz,Oppenheimer:1939ne}
\begin{equation}
	p^{\prime}(r)=-(p(r)+\epsilon(r)) \frac{G_0}{c^2 r^2}\left(M(r)+\frac{4 \pi r^3}{c^2} p(r)\right)\left(1-\frac{2 G_0 M(r)}{c^2 r}\right)^{-1}\,,
\end{equation}
where $M(r)$ denotes the mass contained within a sphere of radius $r$,
\begin{equation}
	M(r)=\frac{4 \pi}{c^2} \int_0^r \epsilon(x) x^2 \mathrm{~d} x\,,
\end{equation}
so that for a star of radius $R_\ast$ the total mass is $M_\ast=M(R_\ast)$. One can see that in the non-relativistic limit one recovers the Newtonian equation of hydrostatic equilibrium. 

\begin{figure}[t]
	\centering 
	\includegraphics[width=0.48\textwidth]{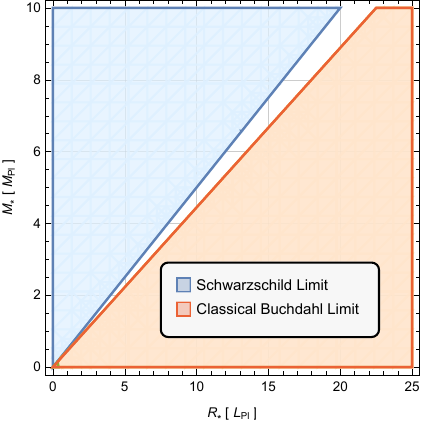}\hfill\includegraphics[width=0.495\textwidth]{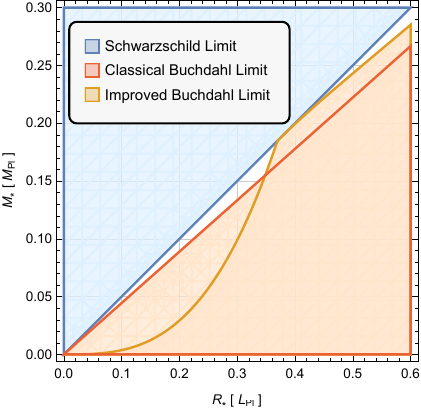}
	\caption{The figures summarize the findings in~\cite{Bonanno:2019ilz} and show the Schwarzschild, classical Buchdahl and improved Buchdahl limits in the $(R_\ast,M_\ast)$ plane, using two different zoom levels. On large scales (see left panel) the classical and quantum-improved cases are indistinguishable. Zooming-in on Planckian scales (see right panel), large deviations become apparent: the improved Buchdahl line approaches the Schwarzschild limit and there exists a critical point at sub-Planckian scales where the two coincides. Below this critical scale, the Buchdahl bound becomes non-linear and the formation of black holes from a collapse becomes more likely due to a larger instability region. 
	Notably, this is consistent with a semi-classical treatment of the problem~\cite{Arrechea:2021xkp}.\label{fig:buchdahl}}
\end{figure}

The importance of the TOV equation in the context of black hole physics lies in the possibility to extract a crucial theoretical bound---known as the Buchdahl limit~\cite{Buchdahl:1959zz}---that regulates the maximal sustainable density of massive stars. Beyond the critical limit, the stellar equilibrium is broken and the star fluid starts collapsing under its own pressure. Too see how this limit emerges classically, let us consider an incompressible fluid with constant proper energy density $\epsilon_0=\rho_0 c^2$. In this case $M(r)=M_\ast r^3/R_\ast^3$ and the TOV equation can be integrated together with the boundary condition $p(R_\ast)=0$, resulting in an expression for the central pressure, 
\begin{equation}
	p(0)=-\frac{3 c^2 M_*\left(3 G_0 M_*+c^2 R_*\left(\sqrt{1-\frac{2 G_0 M_*}{R_* c^2}}-1\right)\right)}{4 \pi R_*^3\left(4 c^2 R_*-9 G_0 M_*\right)}\,.
\end{equation}
The condition that the central pressure $p(0)$ be finite and positive finally yields the classical Buchdahl bound for stellar equilibrium 
\begin{equation}
	M_\ast<\frac{4 c^2}{9 G_0} R_\ast\,.
\end{equation}
If the bound is violated, the configuration is unstable and the object starts collapsing; if no mechanism exists that halts the collapse, the star surface will eventually cross its Schwarzschild radius, thus resulting in a black hole (cf. left panel of Fig.~\ref{fig:buchdahl}).

The question of whether and how the Buchdahl limit are affected by gravitational antiscreening has been tackled in~\cite{Bonanno:2019ilz}. The investigation is based on the RG improvement but follows the idea, first proposed by Markov and Mukhanov~\cite{Markov:1985py}, that the antiscreening character of gravity ought to be included  via an energy-dependent Newton coupling that is introduced as an effective multiplicative coupling between matter and geometry. At the level of the action this boils down to introducing an energy-dependent coupling $\chi(\epsilon)$, so that
\begin{equation}
	\Gamma_0=\frac{1}{16 \pi G_0} \int \mathrm{d}^4 x \sqrt{-g}\,(R+2 \chi(\epsilon) \mathcal{L}_{{matter}})\,.
\end{equation}
Einstein equations are thus modified by an effective energy-momentum tensor
\begin{equation}
	\Lambda_{\mu \nu} \equiv\left(\epsilon \frac{\partial \chi}{\partial \epsilon}+\chi\right) T_{\mu \nu}-\epsilon^2 \frac{\partial \chi}{\partial \epsilon} g_{\mu \nu}\,,
\end{equation}
from which one can read off an effective, energy-dependent version of the Newton coupling and of the cosmological constant
\begin{equation}
	G_{\mathrm{eff}}(\epsilon) \equiv \frac{c^4}{8 \pi} \frac{\partial(\epsilon \chi)}{\partial \epsilon}\,,\qquad \Lambda_{\textrm{eff}}(\epsilon)=-\epsilon^2\frac{\partial\chi}{\partial \epsilon}\,,
\end{equation}
akin to the RG-improved ones. 
Following the same steps as in the classical case and exploiting Eq.~\eqref{eq:running} one can find the quantum-improved version of the central pressure~\cite{Bonanno:2019ilz}. It reads
\begin{equation}
	p(0)=\frac{\epsilon_{\mathrm{Pl}}}{ g_*^{-1}} \frac{\mathcal{N}\left(R_*, M_*\right)}{\mathcal{D}\left(R_*, M_*\right)}\,,
\end{equation}
where $\epsilon_{Pl}$ is a Planckian energy density and
\begin{subequations}
	\begin{align}
		&\mathcal{N}=\left(e ^ { \frac {c ^ { 2 } } { g _ { * } \epsilon _ { \mathrm { Pl } } } \frac { 3 M _ { * } } { 4 \pi R _ { * } ^ { 3 } } - 1  } \right) \left( g_*^{-1} M_* e^{\frac{c^2}{g_* \epsilon_{\mathrm{Pl}}} \frac{3 M_*}{4 \pi R_*^3}}-2 \pi R_*^3 \frac{\epsilon_{\mathrm{Pl}}}{c^2}\left(e^{\frac{c^2}{g_* \epsilon_{\mathrm{Pl}}} \frac{3 M_*}{4 \pi R_*^3}}-1\right)\right) \left(\sqrt{1-\frac{2 G_0 M_*}{R_* c^2}}-1\right)\,, \\
		&\mathcal{D}= g_*^{-1} M_* e^{\frac{c^2}{g_* \epsilon_{\mathrm{Pl}}} \frac{3 M_*}{4 \pi R_*^3}}+2 \pi R_*^3 \frac{\epsilon_{\mathrm{Pl}}}{c^2}\left(e^{\frac{c^2}{g_* \epsilon_{\mathrm{Pl}}} \frac{3 M_*}{4 \pi R_*^3}}-1\right)\left(\sqrt{1-\frac{2 G_0 M_*}{R_* c^2}}-1\right)\, .		
	\end{align}
\end{subequations}
As shown in Fig.~\ref{fig:buchdahl},
the classical and improved Buchdahl bounds coincide at large radii. Nonetheless, substantial deviations become evident in the Planckian region. In there the tilt of the improved Buchdahl limit is smaller than its classical counterpart, so that the quantum Buchdahl line gets arbitrarily close to the Schwarzschild limit. As a consequence, this scenario would predict the existence of 
ultra-compact horizonless objects of Planckian size, whose compactness is beyond the one established by the classical Buchdahl limit. In particular, there exists a critical point at trans-Planckian scales, $R_{cr}\sim 0.37 g_\ast^{-1/2} l_{Pl}$, where the Schwartzschild and improved Buchdahl limit coincide. The classical case, where this critical point is absent, is recovered as the fixed point $g_{\ast}$ is pushed to infinity. Beyond the critical point, the Buchdahl inequality undergoes a major change, indicating a potential transition to a quantum-gravity dominated phase. In such a phase, the function $\mathcal{D}$ turns negative and the Buchdahl limit is determined by the condition $\mathcal{N}<0$, which yields a cubic relation,
\begin{equation}
	R_* \gtrsim 0.27  g_*^{-1}\left(\frac{M_*}{M_{\mathrm{Pl}}}\right)^{\frac{1}{3}} L_{\mathrm{Pl}}\,,
\end{equation}
resembling the scaling relation characterizing ``Planck stars''~\cite{Rovelli:2014cta}. Notably, the general picture is compatible with the results obtained within the semi-classical approach in~\cite{Arrechea:2021xkp}.

\section{Improving the RG improvement: methods and physical results}\label{sect:refined-rg-imp}

The RG improvement provides a straightforward recipe to determine qualitative features of quantum-corrected black-hole (and cosmological) spacetimes---some of which have been summarized in Sect.~\ref{sect:rg-improved-black-holes}. Yet, as remarked in Sect.~\ref{sect:RG-improv-key-idea}, the procedure suffers from a number of problems and ambiguities, making its relation to the asymptotic safety program unsettled. These issues followed several attempts to make the RG improvement procedure more rigorous and to fill the gap between model building and first-principle calculations in quantum gravity. In this section we discuss some of the issues and their proposed partial or complete resolutions.

\subsection{Constraints on the cutoff identification from Bianchi identities}\label{sect:bianchi}

One of the main concerns about the application of the RG improvement is the lack of a clear and systematic recipe to determine the map $k\mapsto k(x)$. The authors of~\cite{Reuter:2004nv,Reuter:2004nx,Babic:2004ev,Domazet:2012tw,Koch:2014joa} have independently shown that some cutoff identifications may be incompatible with diffeomorphism invariance; hence, enforcing the validity of the Bianchi identities could constrain or even completely fix the functional relation~$k(x)$. In this sebsection we will show how this is implemented through some examples. In doing so, we will follow the review~\cite{Platania:2020lqb} closely.

To illustrate how the Bianchi identities can constrain the map~$k(x)$, let us consider the EAA in the Einstein-Hilbert truncation complemented by a matter Lagrangian~$\mathcal{L}_{matter}$,
\begin{equation}
	\Gamma_{k}=\int d^{4}x\sqrt{-g}\;\left(\frac{1}{16\pi G_{k}}(R-2\Lambda_{k})+\mathcal{L}_{matter}\right)\;.\label{eq:qaction}
\end{equation}
We shall neglect the running of the matter couplings in the following. If $k=k(x)$ (in particular, $k(x)\equiv k_{dec}(x)$, cf.~Sect.~\ref{sect:RG-improv-key-idea}), then the classical gravitational field equations will be modified by an effective energy-momentum tensor
\begin{equation}
	\Delta t_{\mu\nu}\equiv G_{k}(\nabla_{\mu}\nabla_{\nu}-g_{\mu\nu}\square)G_{k}^{-1}\,,
\end{equation}
encoding the vacuum polarization effects of the quantum gravitational field~\cite{Reuter:2004nv}. The resulting field equations thus read
\begin{equation}
	G_{\mu\nu}=8\pi G_{k}T_{\mu\nu}-\Lambda_{k}g_{\mu\nu}+\alpha\Delta t_{\mu\nu}\;,\label{eq:modfieldeq}
\end{equation}
where $\alpha=1$ if the RG improvement is performed at the level of the action, whereas $\alpha=0$ if it is employed at the level of field equations.
Enforcing the validity of the Bianchi identities thus yield the ``consistency condition''~\cite{Reuter:2004nv,Reuter:2004nx,Babic:2004ev,Domazet:2012tw,Koch:2014joa}
\begin{align}
	\nabla^{\mu}G_{\mu\nu} & =\left(8\pi G'_{k}T_{\mu\nu}-\Lambda'_{k}\,g_{\mu\nu}\right)\nabla^{\mu}k(x)+8\pi G_{k}\nabla^{\mu}T_{\mu\nu}+\alpha \nabla^{\mu}\Delta t_{\mu\nu}=0\;,\label{eq:bianchiID}
\end{align}
where a prime denotes differentiation with respect to the RG scale $k$. This equation can be exploited to constrain or even determine~$k(x)$. The practical realization of this constraint strongly depends on whether the RG improvement is performed at the level of the action or elsewhere; we shall discuss these two cases separately.

RG improving at the level of the action boils down to setting $\alpha=1$ in Eq.~\eqref{eq:bianchiID}. The modified Bianchi identities thus lead to the conditions~\cite{Koch:2014joa}
\begin{equation}
	\begin{aligned}
	\nabla^{\mu}G_{\mu\nu}  = & +G'_{k}G_{k}^{-1}\left\{ (R_{\mu\nu}-\tfrac{1}{2}Rg_{\mu\nu}-8\pi G_{k}T_{\mu\nu}+\Lambda_{k}g_{\mu\nu})-R_{\mu\nu}\right\} \nabla^{\mu}k(x) \\
	&+\left(8\pi G'_{k}T_{\mu\nu}-\Lambda'_{k}\,g_{\mu\nu}\right)\nabla^{\mu}k(x)+8\pi G_{k}{\nabla^{\mu}T_{\mu\nu}}\label{eq:bianchiID-1}=0\;.
	\end{aligned}
\end{equation}
This relation further simplifies if the energy flow between the gravitational and matter sector is negligible, as in this case the matter energy-momentum tensor $T_{\mu\nu}$ is separately conserved, $\nabla^{\mu}T_{\mu\nu}\approx 0$. Under this condition, the Bianchi identities yield the constraint~\cite{Reuter:2004nv,Domazet:2012tw,Koch:2014joa}
\begin{equation}
	G'_{k}\,R=2(G'_{k}\Lambda_{k}-\Lambda'_{k}G_{k})\;.\label{eq:constraint}
\end{equation}
At this point the specific form of $k(x)$ depends on the running couplings $G_k$ and $\Lambda_k$, which are in turn determined as solutions to the beta functions. As a paradigmatic example, in the fixed point regime  $G_k\sim g_\ast k^{-2}$ and $\Lambda_k\sim \lambda_\ast k^{2}$ so that the above consistency constraint imposes
\begin{equation}
	k^{2}=\frac{R}{4\lambda_{\ast}}\;.\label{eq:cutid}
\end{equation}
A generalization of Eq.~\eqref{eq:constraint} to the case of quadratic gravity was considered in~\cite{Cai:2010zh}, cf.~Eq.~\eqref{eq:gen-cutoffid-R2}. Notably the result~\eqref{eq:cutid} is independent of the specific $f(R)$ truncation, at least when focusing on the fixed-point regime~\cite{Domazet:2012tw,Platania:2019qvo}. The relation~\eqref{eq:cutid} also has important implications in cosmological contexts~\cite{Bonanno:2012jy,Hindmarsh:2012rc,Bonanno:2015fga,Bonanno:2016rpx,Platania:2019qvo,Bonanno:2018gck}, it providing ``a road to modified gravity theories''~\cite{Domazet:2012tw}.

The RG improvement at the level of the field equations is somewhat simpler, as $\alpha=0$ in Eq.~\eqref{eq:bianchiID}. Assuming once again that the matter energy-momentum tensor~$T_{\mu\nu}$ is separately conserved, the consistency condition reads
\begin{equation}
	\left(8\pi G'_{k}T_{\mu\nu}-\Lambda'_{k}\,g_{\mu\nu}\right)\nabla^{\mu}k(x)=0\,,\label{eq:consistency1}
\end{equation}
and depends on the specific matter content of the system. This is to be contrasted with the improvement at the level of the action, where the contribution of the energy-momentum tensor cancels out. A particularly simple while important case is that of a perfect fluid. Indeed, perfect fluids play a role both in cosmology and in gravitational collapse models (see Sect.~\ref{sect:gravitational-collapse}). The energy-momentum tensor of a perfect fluid with energy density~$\rho$ and pressure~$p=w\rho$ is $T_{\mu}^{\nu}=\mathrm{diag}(-\rho,p,p,p)$, and the corresponding consistency condition reduces to
\begin{equation}
	\frac{G_k^\prime}{G_k}\left(\rho+\rho_{\Lambda}(k)\right)+\rho'_{\Lambda}(k)=0\,,
\end{equation}
where we have defined $\rho_{\Lambda}\equiv {\Lambda_k}/{(8\pi G_k)}$. Focusing once again on the fixed-point regime, the above relation suggests that~\cite{Babic:2004ev,Domazet:2010bk},
\begin{equation}
	k^{4}=\left(\frac{8\pi g_{\ast}}{\lambda_{\ast}}\right)\,\rho\,.\label{eq:cutideq}
\end{equation}
A cutoff identification of this type was used in~\cite{Bonanno:2016dyv,Bonanno:2017kta,Bonanno:2017zen} to study the RG-improved gravitational collapse of a massive star into a black hole, as well as in a cosmological context~\cite{Bonanno:2001hi,Bonanno:2001xi,Bonanno:2002zb,Bonanno:2007wg}. Indeed, in cosmology the metric is approximately that of a Friedmann-Lema\^itre-Robertson-Walker spacetime, the energy-density is related to the scale factor via $\rho(a)=\rho_{0}(a(t)/a_{0})^{-3(1+w)}$ (at least assuming the standard conservation equations for $T_{\mu\nu}$, and a power-law behavior for $a(t)$), and $k\propto a(t)^{-\frac{3}{4}(1+w)}$. In turn, $a(t)$ can be written in terms of the cosmological time $t$ or the Hubble constant $H(t)$. It follows that in the proximity of the fixed point, assuming no energy flow between the gravitational and matter sector, all cutoff identifications in terms of time $t$, scale factor $a(t)$, Hubble constant $H(t)$, and matter energy density $\rho(t)$ are physically equivalent (i.e., they produce the same results), as long as they have the correct power.
On the other hand, in regimes where there is a substantial energy exchange between the gravitational and matter sectors, the Bianchi identities are automatically satisfied and thus do not allow to constrain the map $k(x)$.

Finally, as we shall see later, the relation~\eqref{eq:cutideq} is also related to the decoupling mechanism: na\"ively, as $\rho$ also coincides with the Lagrangian density of a perfect fluid, it can act as a physical IR cutoff and thus contribute---together with other IR quantities---to the total decoupling scale $k_{dec}$. The difference between the implementation at the level of the action and at the level of the field equations is resolved when exploiting the decoupling mechanism: the decoupling condition imposes that the decoupling scale be a combination of both the curvature invariants (e.g., the Ricci scalar $R$ in the Einstein-Hilbert truncation) and the matter energy density $\rho$~\cite{Borissova:2022mgd}.

\subsection{Self-consistency and iterative RG improvement}\label{sect:iterativeRGimpro}

The determination of the RG-improved metric $g_{\mu\nu}^{qu}$ starting from the classical (typically Schwarzschild) background $\bar{g}_{\mu\nu}$ involves a scale identification $k=k(x)$ relating it to curvature invariants. Yet, all physical invariants, including the proper distance and the Kretschmann scalar in Eq.~\eqref{eq:sc32}, are built on the classical background metric $\bar{g}_{\mu\nu}={g}_{\mu\nu}^{(0)}$, which is singular and not reliable in the region $r\ll l_{Pl}$, where one needs to determine quantum-gravity induced modifications. The physical invariants constructed using the new, RG-improved metric $g_{\mu\nu}^{qu}={g}_{\mu\nu}^{(1)}$ will generally differ from the classical ones and therefore they would lead to a different functional form for~$k(x)$. Moreover, at variance of the case of scalar QED, the functional expression of $k(x)$--- being built on the spacetime metric---depends explicitly on the coupling that is to be improved (i.e., the Newton coupling $G_0$). Overall, the improvement modifies the spacetime in a way that can impact the improvement itself and thus backreaction effects ought to be accounted for in the procedure. All these points suggest that the RG improvement in gravity should be implemented self-consistently.

A possible approach to find a self-consistent quantum-corrected metric $g_{\mu\nu}^\ast$ is to define a sequence of RG-improvements
\begin{equation}
	g_{\mu\nu}^{cl}=g_{\mu\nu}^{(0)}\quad\rightarrow\quad g_{\mu\nu}^{(1)} \quad\rightarrow\quad \dots \quad\rightarrow\quad g_{\mu\nu}^{(n)} \quad\rightarrow\quad \dots \quad\rightarrow\quad  g_{\mu\nu}^{(\infty)}=g_{\mu\nu}^\ast\,,
\end{equation}
where the metric $g_{\mu\nu}^{(n)}$ at the step $n$ and the corresponding effective Newton coupling~$G_n(r)$ are defined via an IR cutoff which depends on the spacetime at the previous step $k_n[g_{\mu\nu}^{(n-1)}]$. If the procedure converges, it will lead to a self-consistent solution $g_{\mu\nu}^\ast$ which is a fixed point of the iterative procedure: a further RG improvement using the cutoff $k^\ast[g_{\mu\nu}^{(\ast)}]$ would lead to the same RG-improved metric $g_{\mu\nu}^\ast$.

This iterative procedure was first devised and applied in~\cite{Platania:2019kyx}. However, it is worth mentioning that a precursor of these ideas already appeared in~\cite{Pawlowski:2018swz}. Indeed, ~\cite{Pawlowski:2018swz} discussed RG-improved Kerr-(A)dS and Schwarzschild-(A)dS
space-times, using a scale identification based on a partially-improved Kretschmann scalar, where the ``partial'' refers to the omission of the derivatives of the effective Newton coupling $G(r)$ in the expression of the Kretschmann scalar. Using this partially, self-consistent choice, the singularity is not resolved, although it is weaker and some geodesics do not hit it. Other features of the resulting partially self-consistent black holes, such as the horizons and the evaporation process, resemble those of Bonanno-Reuter black holes.

\begin{figure}[t]
	\centering 
	\includegraphics[width=0.65\textwidth]{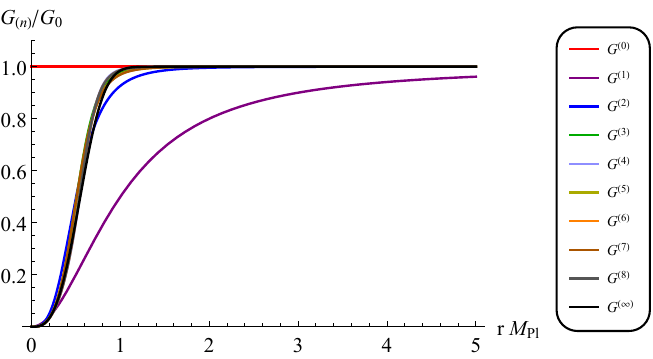}
	\caption{Convergence of the effective Newton coupling $G_n(r)$ (shown in Planck units) in the iterative RG improvement procedure. At the step $n=0$, $G_{n=0}=G_0$. In the first step, the function $G_1(r)$ interpolates between its IR value $G_0$ and zero in the UV. Successive steps display a fast convergence towards the fixed-point solution $G_\infty(r)$.\label{fig:iteration-to-dymni}}
\end{figure}

In~\cite{Platania:2019kyx} the self-consistent RG-improvement procedure was implemented iteratively.
For $n>1$ the metric is determined by the replacement rule
\begin{equation}
	G_{(n)} \rightarrow G_{(n+1)}(r)=\frac{G_0}{1+g_*^{-1} G_0 k_{(n+1)}^2(r)}\,,
\end{equation}
where the cutoff function $k_{(n+1)}$, independent of its functional form (e.g., based on the improved Ricci or Kretschmann scalars), can be written as a functional of the energy-density $\rho_n(r)$ (cf.~Eq.~\eqref{eq:rhop-gen-vaidya}) induced by the spatial variation of $G_{(n)}(r)$ in the previous step, 
\begin{equation}
	k_{(n+1)}^2(r) \equiv \mathcal{K}\left[\rho_{(n)}(r)\right]\,.
\end{equation}
Consequently, the sequence of RG improvements is encoded in a recursive relation,
\begin{equation}
	G_{(n+1)}(r)=\frac{G_0}{1+g_*^{-1} G_0 \mathcal{K}\left[G_{(n)}^{\prime}(r)\right]}\,,
\end{equation}
whose fixed point---reached in the limit $n\to\infty$ under the assumption of convergence---is the solution to the following differential equation
\begin{equation}\label{eq:iterative-rg-equation}
	\mathcal{K}\left[G_{\infty}^{\prime}(r)\right]=g_* \frac{G_0-G_{\infty}(r)}{G_0 G_{\infty}}\,.
\end{equation}
Note that above and in the following we shall adopt the more compact notation $G_{\infty}$ in place of $G_{(\infty)}$ for shortness.
Once a specific cutoff identification $\mathcal{K}$ has been specified, a solution can be found. While the topic of uniquely identifying $k(x)$ is the subject of Sect.~\ref{sect:decoupling-mechanism-solutions}, it is instructive to see how the iterative RG improvement works for a specific choice of $	\mathcal{K}$. Motivated by the form of the improved Ricci and Kretschmann scalars~\cite{Platania:2019kyx}, where one can see that the quantity $G_0\rho$ acts as an effective IR cutoff,~\cite{Platania:2019kyx} set $k^2 \equiv \mathcal{K}[\rho]= G_0 \rho$. The resulting fixed-point equation is the following first-order differential equation,
\begin{equation}
	G_{\infty}(r)=\frac{4 \pi r^2 G_0G_{\infty}(r)}{4 \pi r^2 G_{\infty}(r)+{g_*^{-1} G_0^2 m G_{\infty}^{\prime}(r)}}\,,
\end{equation}
which can be solved exactly and leads to a regular spacetime metric that was put forth two decades earlier by Dymnikova~\cite{Dymnikova:1992ux} (see Fig.~\ref{fig:iteration-to-dymni}). Interestingly, in the classical limit (that can be obtained by pushing the RG fixed point to infinity, $g_\ast\to\infty$) one recovers the Schwarzschild metric, whereas the case of asymptotic freedom $g_\ast\to0$ would require the effective Newton coupling to be everywhere zero.

\subsection{Coordinate dependence and invariant RG improvement}\label{sect:invariant-rg-improvement}

The RG improvement at the level of the solutions implements the replacement $G_0\to G_k$ and the subsequent identification $k\mapsto k(x)$ in the metric coefficients. As argued in~\cite{Held:2021vwd}, a key question is whether this procedure is coordinate independent, i.e., whether the physical properties of the resulting RG-improved spacetime (for instance, its curvature invariants) depend on the initial choice of coordinates. 

While the RG improvement at the level of the action ought to be coordinate independent due to the Lagrangian being a scalar density, a metric $g_{\mu\nu}$ transforms as a $(0,2)$-tensor under coordinate transformations,  implying that the RG improvement at the level of the metric is generally coordinate dependent~\cite{Held:2021vwd}. In the following we review a possible solution to the problem of coordinate dependence---otherwise dubbed ``invariant RG improvement''~\cite{Held:2021vwd}.

To tackle the question of coordinate dependence,~\cite{Held:2021vwd} exploits the characterization of spacetimes of the Petrov and Segre types~\cite{stephani_kramer_maccallum_hoenselaers_herlt_2003} via a complete set of  curvature invariants named Zakhary-McIntosh (ZM) invariants ~\cite{1991JMP....32.3135C,1997GReGr..29..539Z,doi:10.1142/9789812777386_0081}. As these quantities transform as scalars under diffeomorphisms, they can be used to explicitly show that the RG improvement at the level of the metric is coordinate dependent. While we will not go through the details of the proof, which can be found in~\cite{Held:2021vwd}, the intuitive cause of such a coordinate dependence is simply the replacement of a scalar quantity ($G_0\to G[k(r)]$) in an object (the metric coefficients) that does not transform as a scalar under a diffeomorphism transformation. This intuitive understanding also points to a solution to the problem of coordinate dependence: \textit{implementing the RG improvement at the level of the curvature invariants}. Specifically, the invariant RG improvement devised in~\cite{Held:2021vwd} consists in performing the replacement $G_0\to G_k$ in (one of) the functionally independent ZM invariants.

\begin{figure}[t] 
	\centering\includegraphics[width=0.75\textwidth]{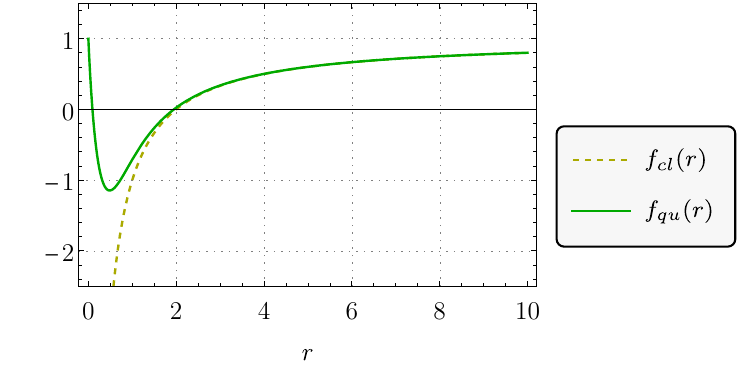}
	\caption{\label{fig:AaronsLapse} Comparison of the classical lapse function (dashed yellow line) with the one in Eq.~\eqref{eq:aaronlapse} (solid green line), which was first computed in~\cite{Held:2021vwd} via the invariant RG improvement. The figure refers to the case of a black hole of Planckian mass, $m=m_{Pl}$, and we set $g_\ast=1$. The critical value to merge and remove the horizons is in this case sub-Planckian, $m_c<m_{Pl}$.}
\end{figure}

In order to make this method more concrete, let us focus on the case of spherically-symmetric black holes and highlight the practical differences between the metric and invariant RG improvements. The metric is in this case diagonal and, assuming $g_{rr}=g_{tt}^{-1}$, it is given by Eq.~\eqref{eq:line-element}. A  Schwarzschild black hole is characterized by the classical lapse function
\begin{equation}
	f_{cl}(r)=1-\frac{2mG_0}{r}\,,
\end{equation}
and by only one independent ZM invariant 
\begin{equation}
	\left(\frac{\mathcal{K}_1^{cl}}{48}\right)^3=\left(\frac{\mathcal{K}_3^{cl}}{96}\right)^2=\left(\frac{G_0 M}{r^3}\right)^6\,,\quad \mathcal{K}_{i\neq 1,3}^{cl}=0\,.
\end{equation}
At this point the standard RG improvement at the level of the metric follows the rule
\begin{equation}
	f_{cl}(r)\quad\underset{\text{Metric RG impr.}}{\longrightarrow}\quad f_{qu}=1-\frac{2mG[k(r)]}{r}\,,
\end{equation}
for some choice of $k=k(r)$, while the invariant RG improvement is defined by
\begin{equation}
	\mathcal{K}_1^{cl}(r)\quad\underset{\text{Invariant RG impr.}}{\longrightarrow}\quad \mathcal{K}_1^{qu}=48\left(\frac{mG[k(r)]}{r^3}\right)^2\,.
\end{equation}
Within this scheme, the new, RG-improved lapse function is defined by the differential equation
\begin{equation}
	\mathcal{K}_1=\frac{(r^2 f_{qu}^{\prime\prime}-2rf_{qu}^{\prime}+2f_{qu}-2)^2}{3r^4}\equiv \mathcal{K}_1^{qu}\,.
\end{equation}
The result at this point depends on the cutoff function $k(r)$. Identifying for instance $k^4\sim \mathcal{K}_1\equiv K $ yields the effective RG-improved metric (see Fig.~\ref{fig:AaronsLapse})
\begin{equation}\label{eq:aaronlapse}
	\begin{aligned}
		f_{qu}(r)= 1&-\frac{2 r \ell}{g_\ast^{-1}l_{Pl}^2}\left[\sqrt{3} \pi+ 2\sqrt{3} \arctan \left(\frac{1-2 r / \ell}{\sqrt{3}}\right)+\log \left(\ell^2-r \ell+r^2\right)-2 \log (\ell+r)\right] \\
		&+\frac{4 r^2}{g_\ast^{-1}l_{Pl}^2}\left[\log \left(\ell^3+r^3\right)-3 \log (r)\right] \, ,
	\end{aligned}
\end{equation}
with $\ell=\sqrt[3]{g_\ast^{-1} l_{Pl}^4 m}$. An asymptotic expansion of this lapse function shows that this metric reproduces the Schwarzschild spacetime for $r\to\infty$, together with additional corrections. In the opposite limit, $r\to0$, where deviations due to quantum gravity are stronger, the lapse function depends linearly on the radial coordinate. Quantum gravity effects thus make the singularity weaker, but do not resolve it. A possible reason, as argued in~\cite{Held:2021vwd}, is that the invariant RG improvement ought to be combined with the iterative procedure described in the previous section; this combination would allow to determine an effective self-consistent metric via a coordinate-independent procedure. On top of this, a clear recipe to fix the functional relation $k(x)$ uniquely would make the RG improvement procedure overall more consistent and rigorous. The latter issue will be the focus of the next subsection.

\subsection{Effective solutions from the decoupling mechanism}\label{sect:decoupling-mechanism-solutions}

This subsection focuses on the findings in~\cite{Borissova:2022mgd}, which for the first time exploited the decoupling mechanism (cf. Sect.~\ref{sect:RG-improv-key-idea}) to uniquely determine the functional relation~$k(x)$ grounded on RG considerations. The resulting framework was then applied, in combination with the iterative RG improvement (cf. Sect.~\ref{sect:iterativeRGimpro}) to investigate the dynamics of self-consistent quantum-corrected black holes from formation to evaporation, within the same VKP collapse model used in Sect.~\ref{sect:gravitational-collapse}.

The first step to exploit the decoupling mechanism~\cite{Reuter:2003ca} is to find a well-defined mathematical expression for the decoupling condition. Its determination, as briefly discussed in Sect.~\ref{sect:RG-improv-key-idea}, could yield a short-cut from a truncated version of the EAA $\Gamma_k$ to the effective action $\Gamma_0$ (or its solutions, to some extent), typically within a larger truncation than the original one. Inasmuch as the decoupling occurs when physical IR scales in the effective action (if any) overcome the artificial regulator $\mathcal{R}_k\sim k^2$, the decoupling condition can be mathematically characterized at the level of the inverse propagator. The latter has the following schematic structure
\begin{equation}
	\Gamma^{(2)}_k+\mathcal{R}_k= c\,( p^2+A_k[\Phi]+\tilde{\mathcal{R}}_k)\,,
\end{equation}
where $c$ is a constant, $\Phi$ is the set of fields considered, and $A_k[\Phi]\equiv  \Gamma^{(2)}_k/c -p^2 $ and $\mathcal{R}_k\equiv c \, \tilde{\mathcal{R}}_k$. The decoupling condition thus
reads~\cite{Borissova:2022mgd}
\begin{equation}\label{eq:dec-condition}
	\tilde{\mathcal{R}}_{k_{dec}} \approx A_{k_{dec}}[\Phi] \,,
\end{equation}
and provides an implicit way to uniquely identify $k$ as the decoupling scale $k_{dec}$. Notably, even in the presence of multiple physical IR scales (e.g., masses, curvature invariants, and matter energy density), the decoupling scale is given by a precise combination of all of them. This is not surprising, as the dynamics of a system typically depends on all its components; similarly, all physical scales should
contribute---if perhaps with different weights---to the determination of the sought-after decoupling scale.

It is pedagogical to see explicitly how this works in massless scalar QED~\cite{Platania:2018eka}. For this case, $\Gamma_k^{(2)}\approx p^2+\lambda_k\phi^2+\tilde{\mathcal{R}}_k+\dots$, with $\lambda_k\approx\log k$ being the coupling in front of the $\phi^4$ interaction term. Choosing a mass-type regulator, $\tilde{\mathcal{R}}_k\sim k^2$, the decoupling condition yields---to leading-order---the Coleman-Weinberg potential $\phi^4\log \phi$.

\begin{figure}[t] 
	\centering\includegraphics[width=0.65\textwidth]{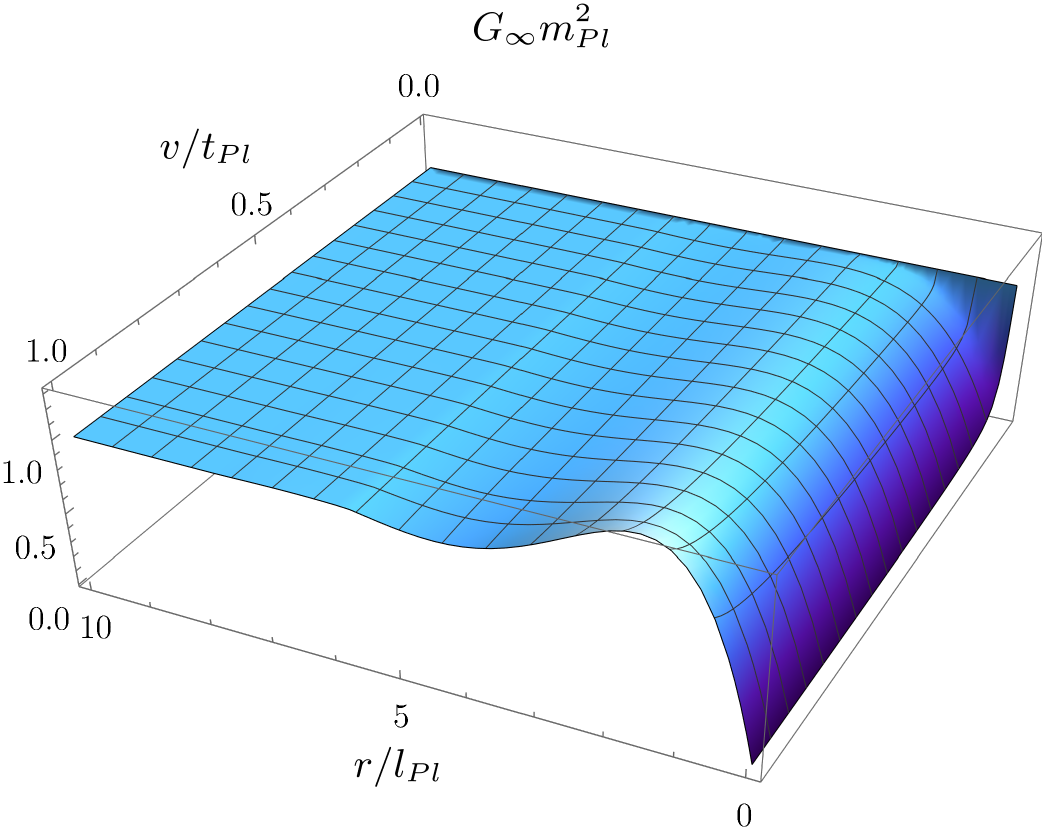}
	\caption{\label{fig:BP-1} Numerical solution to the partial differential equation~\eqref{eq: G infinity differential equation} self-consistently defining the effective Newton coupling associated with a dynamical quantum-corrected spacetime stemming from the decoupling mechanism~\cite{Borissova:2022mgd}. In order to solve Eq.~\eqref{eq: G infinity differential equation},~\cite{Borissova:2022mgd} used the boundary and initial conditions ~$G_\infty(r,v_0) =  G_\infty(r_{max},v) = G_0$. A final boundary condition is imposed by requiring the solution to match the solution close to the would-be singularity, as this can be studied analytically~\cite{Borissova:2022mgd}. By construction, the effective Newton coupling reproduces the observed value of Newton constant at early times and at large distances. For $v>0$ and during the entire duration of the collapse, the advanced time dependence of the effective Newton coupling is $\sim v^{-1}$. Once the infusion of radiation from the nearby massive star is over, the collapse ends and the spacetime is described by an effective Newton coupling which is monotonic with respect to the radial coordinate and smoothly connects the limiting values $G_0$, in the IR, and $0$, in the UV. In addition, this dynamical black hole spacetime is characterized by damped oscillations along the radial direction, reminiscent of higher-derivative operators with specific non-local form factors~\cite{Zhang:2014bea}.}
\end{figure}

Let us now turn on gravity and exploit the decoupling mechanism to study the quantum-corrected gravitational collapse in a VKP model. Within the Einstein-Hilbert truncation the EAA reads
\begin{equation}\label{eq: effective action}
	\Gamma_k =  \int \dd[d]x \sqrt{g}\qty(\frac{1}{16\pi G_k}\qty(2\Lambda_k-R) + \mathcal{L}_m) \,,
\end{equation}
where $G_k$ and $\Lambda_k$ are the RG-scale dependent Newton coupling and cosmological constant, $d$ is the number of spacetime dimensions, and $\mathcal{L}_m$ is the Lagrangian of a pressureless perfect fluid (radiation)~\cite{Ray:1972},
\begin{equation}\label{eq: matter Lagrangian}
	\mathcal{L}_m = \mu(r,v) \,,
\end{equation}
as required by the VKP model. Complementing the gravitational EAA with a harmonic gauge fixing and the standard Faddeev-Popov ghost action, setting $d=4$, and restricting to a mass-type regulator, the decoupling condition reads
\begin{equation}\label{eq:dec-cond}
	k_{dec}^2\equiv G_{k_{dec}} \mu +\frac{2}{3} R \,,
\end{equation}
where $G_{k_{dec}}$ is given by Eq.~\eqref{eq:running}, as usual, with $k\equiv k_{dec}$. One can straightforwardly see that replacing this condition in the EAA introduces higher-derivative operators in the effective action, as expected. We note at this point that $R$ can always be written in terms of the gravity-induced energy density and pressure~\eqref{eq:rhop-gen-vaidya}; in turn $\mu$, $\rho$ and $p$ depend on the effective Newton coupling $G(r,v)$ and its partial derivatives. Therefore, one can straightforwardly apply the differential equation defining the fixed point of the iterative RG improvement---Eq.~\eqref{eq:iterative-rg-equation}---with the cutoff identification $\mathcal{K}$ dictated by the decoupling condition~\eqref{eq:dec-cond}. 
The differential equation determining the self-consistent effective Newton coupling is~\cite{Borissova:2022mgd}
\begin{equation}\label{eq: G infinity differential equation}
	G_\infty = \frac{G_0}{1+ g_\ast^{-1}G_0 G_\infty \qty(\mu_\infty  + \frac{2}{3}  16 \pi \qty(\rho_\infty - p_\infty))}\,,
\end{equation}
where $\mu_\infty$, $\rho_\infty$ and $p_\infty$ are given by
\begin{equation}\label{eq:explicitdependence}
	\mu_{\infty} = \frac{\dot{m}(v)}{4\pi G_{\infty} r^2}+\frac{m(v)\dot{G}_{\infty}}{4\pi G_{\infty} r^2}\,, \quad\quad \rho_{\infty} = \frac{m(v)\,G' _{\infty}}{4\pi G_{\infty} r^2}\,, \quad\quad p_{\infty} = -\frac{m(v)\,G'' _{\infty}}{8\pi G_{\infty} r}\,,
\end{equation}
and, as usual, primes and dots denote spatial and advanced time derivatives, respectively. Solutions to this differential equation for a given collapse model, specified by the mass function $m(v)$, correspond to quantum-corrected dynamical spacetimes with lapse function
\begin{equation}
	f_{qu}(r,v)\equiv f_\infty(r,v)=1-\frac{2\,m(v)\,G_{\infty}(r,v)}{r}\,,
\end{equation}
describing the spacetime dynamics from formation to evaporation. In particular,~\cite{Borissova:2022mgd} studied in detail solutions to the defining differential equation~\eqref{eq: G infinity differential equation} for a VKP model, both analytically, in different limiting regimes, and numerically. The full numerical solution is displayed in Fig.~\ref{fig:BP-1}.
and is derived by imposing that the observed Newton coupling is recovered at early times (before the collapse starts, for any $r$) and at large distances,~$G_\infty(r,v_0) =  G_\infty(r_{max},v) = G_0$. The remaining boundary condition is set by determining the asymptotic behavior of the solution analytically,~\cite{Borissova:2022mgd}.
By construction, the effective Newton coupling reproduces the observed value of Newton constant at early times and at large distances. During the collapse the effective Newton coupling decays as $\sim v^{-1}$. At the end of the collapse the effective Newton coupling converges to a function that connects smoothly the IR regime (large radii), where $G_\infty(r)\to G_0$, and the UV fixed-point scaling ($r\to0$), where the effective Newton coupling vanishes. While all these features were expected from previous studies, quantum-corrected black holes stemming from the decoupling mechanism feature an additional striking feature reminiscent of higher-derivative operators with specific non-local (exponential) form factors~\cite{Zhang:2014bea}: damped oscillations along the radial direction. The presence of non-trivial form factors is crucial to explain the oscillations. Indeed, black holes in quadratic gravity lead to free oscillations provided that a specific sign for the Weyl square term $C^2$ in the action is used~\cite{Bonanno:2013dja,Bonanno:2019pmm,Bonanno:2019rsq}, while the damping can be obtained from the presence of exponential form factors~\cite{Zhang:2014bea}. This is in line with the expectation that the decoupling mechanism ought to grant access to higher-derivative operators that were not initially considered in the original truncation for the EAA---the Einstein-Hilbert one in the case of~\cite{Borissova:2022mgd}. 

\begin{figure}[t] 
	\hspace{-0.3cm}\includegraphics[width=0.52\textwidth]{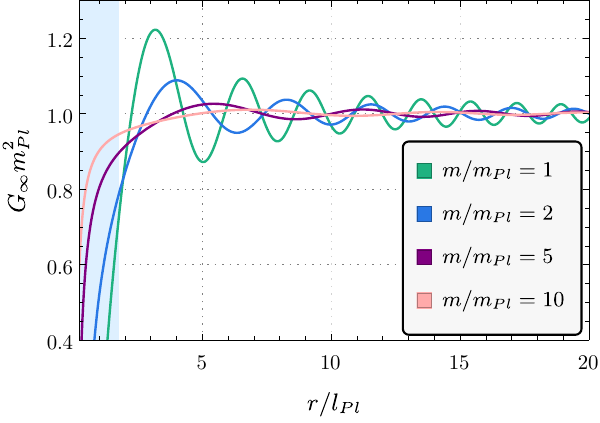}\hspace{-0.1cm}\includegraphics[width=0.51\textwidth]{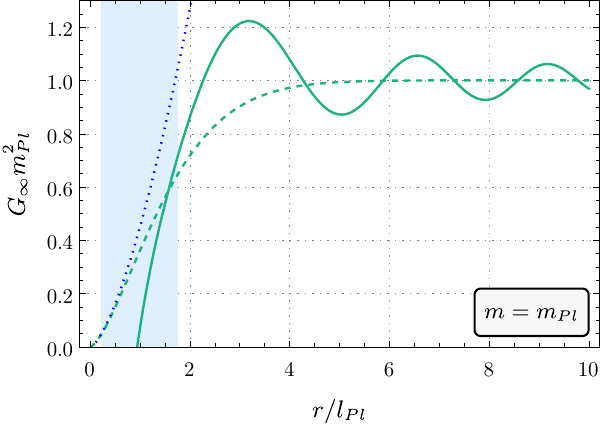}
	\caption{\label{fig:BP-2} Static limit of the dynamical effective Newton coupling at the end of the collapse. The figure on the left depicts the static solution at large distances, for different values of the black hole mass (corresponding to differently-colored lines). It highlights that oscillations are present also in the static limit, with an amplitude that depends on the black hole mass. The figure on the right refers to Planckian black holes only, and compares three approximations: the $\sim r^{3/2}$ scaling close to the singularity (dotted line, Eq.~\eqref{eq: G infinity fixed point static solution}), the large-distance solution described by Airy functions (solid line, Eq.~\eqref{eq: Airy F}), and the interpolating function which connects these two regimes but neglects oscillations (dashed line, Eq.~\eqref{eq: approximate static solution}). The blue region denotes the transition zone between the small and large radii limit: in this region the two limiting solutions (dotted and solid lines) cannot be trusted~\cite{Borissova:2022mgd}.}
\end{figure}

One may speculate that exponential form factors would lead to exponential lapse functions; this expectation can be partially verified by studying the static limit of the dynamical solution, corresponding to the static configuration $m(v)\to m$ at the end of the collapse. This can be done analytically, within three complementary approximations: the short-distance regime close to the classical singularity, the large-distance limit, and an interpolation between the two that neglects the damped oscillations. The first limiting regime is reached by approximating $G_k\sim g_{\ast} k^{-2}$ in the derivation of Eq.~\eqref{eq: G infinity differential equation}. This is tantamount to neglecting the 1 in the denominator of Eq.~\eqref{eq: G infinity differential equation}. The resulting equation
\begin{equation}\label{eq: G infinity differential equation static fixed-point}
	G_0 g_\ast^{-1} m  \qty(4 r G_\infty '' + 8 G_\infty ')G_\infty  - 3 G_0 r^2 = 0\,,
\end{equation}
can be solved by parameterizing $G_{\infty}(r)\sim C\,r^n$ close to $r=0$. With this strategy one finds the leading-order behavior of $G(r)$ close to the classical singularity,
\begin{equation}\label{eq: G infinity fixed point static solution}
	G_\infty(r)  =  \frac{1}{\sqrt{5 g_\ast^{-1} m}}r^{3/2}\,.
\end{equation}
This scaling (dotted line in the right panel of Fig.~\ref{fig:BP-2}) is sufficient to make $G_\infty(r)$ vanish in the UV, but the decay with the radial coordinate $r$ is not fast enough to resolve the singularity (cf.~Sect.~\ref{sect:cosmolo-const-role}). This is nonetheless expected from general considerations from the gravitational collapse (cf.~Sect.~\ref{sect:gravitational-collapse}). Next one could consider the opposite limit, $r\to\infty$, where the differential equation~\eqref{eq: G infinity differential equation} reduces to
\begin{equation}\label{eq: G infinity differential equation static}
	\qty(G_0 g_\ast^{-1} m \qty(4 r G_\infty '' + 8  G_\infty ')   + 3 r^2)G_\infty - 3 G_0  r^2 = 0\,.
\end{equation}
In order to solve this equation one can make the ansatz~\cite{Borissova:2022mgd}
 \begin{equation}\label{eq: G infinity large r ansatz}
	G_\infty(r) = G_0\,\qty(1 - \frac{F(r)}{r})\,,
\end{equation}
with $\abs{F(r)/r}\ll 1$ asymptotically, and  $\abs{F(r)/r}\to 0$ as $r\to \infty$. Inserting the ansatz~\eqref{eq: G infinity large r ansatz} into the differential equation~\eqref{eq: G infinity differential equation static} and solving with respect to the function $F(r)$ yields the solution
\begin{equation}\label{eq: Airy F}
	F(r) = \mathfrak{Re} \big[c_1 \text{Ai}(a(m,g_\ast) r) + c_2 \text{Bi}(a(m,g_\ast)r)\big]\,,   
\end{equation}
where $a(m,g_\ast) = 2^{-2/3}3^{1/3}(-G_0^2 m g_\ast^{-1})^{-1/3} $ and $c_i \propto 1/m$ on dimensional grounds. The corresponding effective Newton coupling is displayed in Fig.~\ref{fig:BP-2}.
Finally, neglecting the damped oscillations, one can determine an interpolating function between the power-law scaling~\eqref{eq: G infinity fixed point static solution} at small radii and the Newton constant at large distances. This function reads
\begin{equation}\label{eq: approximate static solution}
	G_\infty(r) = G_0\qty(1 - e^{-\frac{r^{3/2}}{\sqrt{5 g_\ast^{-1} 2mG_0/2}\,l_{Pl}}})\,, 
\end{equation}
and is displayed in right panel of Fig.~\ref{fig:BP-2}, together with the other two approximate expressions of $G_\infty(r)$ derived above~\cite{Borissova:2022mgd}. The interpolating function is an exponential, which is expected from the self-consistent implementation of the RG improvement (cf. Sect.~\ref{sect:iterativeRGimpro}) and from the tentative relation with the exponential non-local form factors potentially responsible for the damped oscillations of the lapse function. The exponential nature of the lapse function is also highly desirable since polynomial asymptotic scalings are not always compatible with the principle of least action~\cite{Knorr:2022kqp}. We shall come back to this topic Sect.~\ref{sect:least-action-principle-bh}.

\section{Towards black holes from first principles}\label{sect:bh-from-frg}

In this section we will summarize four independent investigations constraining aspects of quantum black holes, within and beyond asymptotic safety, hinging on first-principle calculations or considerations involving the FRG, the path integral, or the effective action. We shall present the corresponding findings in four subsections, following a chronological order.

\subsection{State counting and entanglement entropy in asymptotic safety}\label{sect:martin-entropy}

In this subsection we review key findings on the topic of black hole entropy in asymptotically safe gravity. See also~\cite{Falls:2012nd,Zhang:2018xzj,Chen:2022xjk} for complementary works on the wider subject of black hole thermodynamics in asymptotic safety.

\paragraph{{\bf Microstate counting}~\cite{Becker:2012js,Becker:2012jx,Koch:2013owa}}

The scope of~\cite{Becker:2012js,Becker:2012jx,Koch:2013owa} was to provide an explanation for the area-scaling of the Bekenstein-Hawking entropy in terms of microscopic gravitational degrees of freedom. Becker and Reuter did so for Schwarzschild black holes in~\cite{Becker:2012js,Becker:2012jx} by proposing a state-counting formula based on the EAA~$\Gamma_k$. Their calculation was subsequently applied to other types of black holes in~\cite{Koch:2013owa}.

The derivation is based on a (Euclidean) path-integral, on-shell representation of the EAA,
\begin{equation}
	\mathbb{Z}_k \equiv e^{-\Gamma_k\left[0, \bar{g}_k^{\text {sf }}\right]}=\int \mathcal{D} \hat{\Phi} e^{-\tilde{S}\left[\hat{\Phi}, \bar{g}_k^{\text {sf }}\right]} e^{-\Delta_k S[\hat{\Phi}]}\,
\end{equation}
where the integration is restricted to fluctuations $\hat{\Phi}$ whose average is zero, $\langle\hat{\Phi}\rangle=0$. Moreover, $\tilde{S}$ stands for the sum of the bare, gauge fixing, and ghost actions,
$\Delta_k S[\hat{\Phi}]\sim\int d^dx \sqrt{-\bar{g}}\,\hat{\Phi}R_k\hat{\Phi}$ is the 
regulator which also appears in Eq.~\eqref{eq:flow-eq}. Moreover, $\bar{g}_k^{\text {sf}}$ is a so-called ``self-consistent background'', i.e., a solution to the scale-dependent field equations defined by $\Gamma_k$,
\begin{equation}\label{eq:self-cons-background}
	\left.\frac{\delta \Gamma_k[h ; \bar{g}]}{\delta h_{\mu \nu}} \right|_{h=0,\bar{g}=\bar{g}_k^{\text {sc}}}=0\,,
\end{equation}
that is also used as expansion point in the background field method, $g=\bar{g}_k^{\text{sc}}+h$.

The analysis can then be made concrete by specifying the bare action and by picking one of its specific solutions as a background field. Focusing on the case of the Einstein-Hilbert truncation with vanishing cosmological constant, solutions are Ricci flat spacetimes, and (Euclidean) Schwarzschild black holes constitute a prime example. Accordingly,  \cite{Becker:2012js,Becker:2012jx,Koch:2013owa} fixed the self-consistent background to be a Euclidean Schwarzschild solution, with $r\in[r_s,\infty)$ and $t\in[0,\beta=4\pi r_s\equiv T^{-1}]$.

As the Euclidean Schwarzschild solution is Ricci flat, the bulk action vanishes on shell, and the only contribution to the partition function comes from the boundary terms, so that the entropy of modes with $p^2>k^2$ is given by
\begin{equation}\label{eq:entropy-boundary}
	\mathcal{S}=-\ln \mathbb{Z}_k=-\frac{1}{8 \pi G_k^{ \partial}} \int_{\partial \mathcal{M}} \mathrm{d}^3 x \sqrt{\bar{H}}\left(\bar{K}-\bar{K}_0\right)+\cdots=\frac{\beta r_s}{4G_k^{\partial}}+\dots\,
\end{equation}
where $\bar{K}$ is the extrinsic curvature on the induced spatial background $\bar{H}$, and $G_k^{\partial}$ is the boundary Newton coupling---defined as the Newton coupling appearing in front of the boundary terms. In the limit where all fluctuating modes are integrated out, $k\to0$, and provided that one can find a suitable trajectory with $G_k^{\partial}\to G_0$, the above expression yields the standard form for Bekenstein-Hawking area law.

\paragraph{{\bf Finiteness of entanglement entropy}~\cite{Pagani:2018mke}}

In~\cite{Pagani:2018mke} the authors focused on another aspect of entanglement entropies, namely, their infamous UV quadratic divergences.

Let us consider a quantum system with Hilbert space given by the direct product of two Hilbert spaces, $\mathcal{H}= \mathcal{H}_A \otimes \mathcal{H}_B$. If the system is in the pure state $|\psi\rangle$, its total density matrix is $\rho=|\psi\rangle\langle\psi|$ while the reduced one for the subsystem A is obtained by tracing over $\mathcal{H}_B$. The entaglement entropy between the two subsystems then equals the von Neumann entropy 
\begin{equation}
	\mathcal{S}=-\mathrm{Tr}[\rho_A \log \rho_A]\equiv - \lim_{n\to1} \frac{\partial}{\partial n} \mathrm{Tr}[\rho_A^n]\,.
\end{equation}
Concretely, its evaluation makes use of the replica trick, which boils down to a relation between the entanglement entropy and certain partition functions. To illustrate how this works, let us consider a free quantum field on a Minkowski spacetime, and let us introduce a spatial surface $\Sigma$ at $x_\mu=0$, such that at $t=0$ the field's degrees of freedom are located either at $x<0$ (subsystem A) or at $x>0$ (subsystem B). Then, the entanglement entropy between the two is given by
\begin{equation}\label{eq:ent-entropy}
	\mathcal{S}=\left.\left[1+2 \pi \frac{d}{d \delta}\right] \log Z_\delta\right|_{\delta=0}\propto {A_\Sigma[\bar{g}]}\,\Lambda_{UV}^2\,,
\end{equation}
where $Z_\delta$ is the partition function of the quantum field on a conical spacetime with deficit angle $\delta$, $A_\Sigma[\bar{g}]$ is the proper area of the surface $\Sigma$ with respect to the background metric $\bar{g}$, and $\Lambda_{UV}$ is a UV momentum cutoff which is to be removed. It is thus clear that the entanglement entropy diverges quadratically as $\Lambda_{UV}\to\infty$.
The question tackled in~\cite{Pagani:2018mke} is whether these divergences are removed by switching quantum gravity fluctuations on, as in this case the rigid background would be replaced by a dynamical,  curved, fluctuating spacetime.

According to~\cite{Pagani:2018mke}, the key to understand quantum-gravity-induced modifications to Eq.~\eqref{eq:ent-entropy} is the observation that in a quantum-gravitational setting $\bar{g}$ is replaced by a self-consistent, scale-dependent background, cf.~Eq.~\eqref{eq:self-cons-background}. Here in particular the cosmological constant plays a crucial role. Focusing on the case of the Einstein-Hilbert truncation, Eq.~\eqref{eq:self-cons-background} yields
\begin{equation}
	R_\nu^\mu\left(\bar{g}_k^{\mathrm{sc}}\right)-\frac{1}{2} \delta_\nu^\mu R\left(\bar{g}_k^{\mathrm{sc}}\right)+\Lambda_k \delta_\nu^\mu=0\,.
\end{equation}
Solutions to this equation are such that $\Lambda_k \bar{g}_k^{\mathrm{sc}}=\text{const}$ and thus, in particular,
\begin{equation}
	\left(\bar{g}_k^{\mathrm{sc}}\right)_{\alpha \beta}=\frac{\Lambda_\mu}{\Lambda_k}\left(\bar{g}_\mu^{\mathrm{sc}}\right)_{\alpha \beta}=\frac{\mu^2 \lambda_\mu}{k^2 \lambda_k}\left(\bar{g}_\mu^{\mathrm{sc}}\right)_{\alpha \beta} \,.
\end{equation}
where $\mu$ is an arbitrary scale. The quadratic dependence on the momenta $k$ and $\mu$ can at this point be exploited to redefine the metric in terms of its dimensionless counterpart, that we shall call $\tilde{\bar{g}}$,
\begin{equation}\label{eq:sc-metric-rescaling}
	\left(\tilde{\bar{g}}_k^{\mathrm{sc}}\right)_{\alpha \beta}=\frac{\lambda_\mu}{ \lambda_k}\left(\tilde{\bar{g}}_\mu^{\mathrm{sc}}\right)_{\alpha \beta} \,.
\end{equation}
Thanks to the asymptotic safety condition, for which $\lambda_k$ approaches the constant value~$\lambda_\ast$ asymptotically, the above equation is well defined in the limit $k\to\infty$. Together with the observation that the proper area function is linear, $A_\Sigma[c\bar{g}]=c A_\Sigma[\bar{g}]$, one finds that in asymptotic safety the limit~\cite{Pagani:2018mke}
\begin{equation}\label{eq:ent-entropy-as}
	\mathcal{S}\propto \lim_{\Lambda_{UV}\to \infty} {A_\Sigma[\bar{g}^{sc}_{\Lambda_{UV}}]}\,\Lambda_{UV}^2=A_\Sigma[\tilde{\bar{g}}^{sc}_{\ast}]\,,
\end{equation}
i.e., the entanglement entropy is finite. Moreover, by means of Eq.~\eqref{eq:sc-metric-rescaling}, the above expression can be re-written as
\begin{equation}
	\mathcal{S}\propto \frac{\Lambda_\mu}{\lambda_\ast} A_\Sigma[\tilde{\bar{g}}^{sc}_{\mu}]\,,
\end{equation}
Keeping in mind that the total entanglement entropy ought to be independent of the normalization scale $\mu$, one may fix it arbitrarily. Specifically, setting $\mu=m_{Pl}={G_0^{-1/2}}$ one finds an expression that is close to the Bekenstein-Hawking formula and resembles the result~\eqref{eq:entropy-boundary} discussed above and first found in~\cite{Becker:2012js}.

\subsection{Towards a dressed Newtonian potential within asymptotic safety}\label{sect:newpotential}

The metric of Schwarzschild or structurally similar modified black holes can be written in terms of a lapse function
\begin{equation}
	f(r)=1-2\Phi(r)\,,
\end{equation}
where $\Phi(r)$ is sometimes dubbed ``Newtonian potential''. It is important to remark that this a slight abuse of notation, since the identification of $\Phi(r)$ with the classical Newtonian potential---defined as the potential exerted by a massive point source on a test particle (hence, non-interacting and non-massive, so as not to impact the spacetime)---only holds asymptotically, in the weak field regime. 

In a quantum gravity context this pseudo-Newtonian potential ought to be determined by solving the quantum field equations~\eqref{eq:full-eom}. In turn, this would require the computation of a sufficiently accurate approximation to the effective action. While such an expression has not been computed yet, one may get an intuition on the form of the Newtonian potential by calculating the graviton-mediated $2\to2$ scattering amplitude of two scalar fields minimally coupled to gravity, in the static limit. At one loop, this reproduces the well-known result by Donoghue on the leading-order corrections to the Newtonian potential~\cite{Donoghue:1993eb}. First steps towards the extension of Donoghue's calculation beyond one-loop\footnote{It is worth mentioning that the one-loop results in~\cite{Donoghue:1993eb} have not been reproduced yet via full-fledged FRG computations. See however~\cite{Satz:2010uu} for first steps in this direction.} and within asymptotic safety have been taken in~\cite{Bosma:2019aiu}. The scope of this subsection is to summarize the work of~\cite{Bosma:2019aiu} and its conclusions.

Prior to starting, it is important to remark that the identification of the Newtonian potential with the one stemming from a graviton-mediated $2\to2$ scattering between matter or gauge fields, which we shall denote by $V(r)$, might not hold in general. On the formal side, indeed, the first is the potential exerted by a massive static source on a test particle, while the second is the potential between two massive and interacting fields. On the practical side, there exists already a clear counterexample where this identification does not work~\cite{Ferrero:2021lhd}: when introducing a cosmological constant, the Newtonian potential is a simple asymptotically de Sitter spacetime, whereas the potential from a $2\to2$ scattering amplitude is characterized by a de Sitter horizon and vanishes beyond it. Given the argument as well as the counterexample in~\cite{Ferrero:2021lhd}, it is unlikely that the two potentials will coincide in a full-fledged quantum gravity computation. Yet, determining $V(r)$ is important on several levels, including the comparison with the EFT results~\cite{Donoghue:1993eb}, and could provide some hints on singularity resolution in asymptotic safety~\cite{Bosma:2019aiu}. 

The work in~\cite{Bosma:2019aiu} considers the effective gravitational scattering of two scalar fields minimally coupled to gravity. The calculation is performed in the static approximation, where both scalar fields are infinitely massive, $m_i\to\infty$. In this limit the potential between the two scalar masses reads~\cite{Donoghue:1993eb,Bosma:2019aiu}
\begin{equation}\label{eq:potentialfromM}
	V(r)=-\frac{1}{2 m_1} \frac{1}{2 m_2} \int \frac{d^3 \mathbf{q}}{(2 \pi)^3} e^{i \mathbf{q} \cdot \mathbf{r}} \mathcal{M}(\mathbf{q}^2)\,,
\end{equation}
where $\mathcal{M}$ is the scattering amplitude of two scalars into two scalars, mediated by one graviton line,
\begin{equation}
	\mathcal{M}(\mathbf{q}^2)=16\pi G_0 m_1^2 m_2^2\, \mathcal{G}(\mathbf{q}^2)\,,
\end{equation}
with $\mathcal{G}(\mathbf{q}^2)$ being the scalar part of the (dressed) graviton propagator in the non-relativistic limit $q\to (0,\mathbf{q})$. For small momenta, one may neglect higher-derivative operators in the effective action, and the scaling of the propagator reduces to the Einstein-Hilbert one, $\mathcal{G}\propto 1/q^2$. Correspondingly, on large distance scales the leading-order scaling of the exact potential $V_{qu}(r)$ is well approximated by its classical counterpart, $V_{cl}(r)\propto 1/r$; in this subsection we explain the steps taken in~\cite{Bosma:2019aiu} to derive a quantum version of $V_{cl}$ within asymptotic safety. The technical difficulties involved in the computation of $V_{qu}$ require however a number of approximations, on top of the static limit, that we detail in the following.

$V_{qu}$ is determined by the full momentum dependence of the dressed propagator. In turn, this is encoded in the form factors $g_R({\Box})$ and $g_C({\Box})$ in the effective action~\eqref{eq:eff-action}. Focusing on the contribution from the transverse-traceless (TT) spin-2 mode, the scalar part of the dressed TT propagator reads
\begin{equation}
	\mathcal{G}^{TT}(q^2)=q^{-2}\left(1+2\,q^2g_C(q^2)\right)^{-1}\,.
\end{equation}
In particular, the function $g_C(q^2)$ is to be computed from first principles, e.g., either by solving the path integral or by determining the IR limit, $k\to0$, of its flowing counterpart $g_{C,k}(q^2)$. The latter strategy involves inserting the ansatz~\eqref{eq:EAA-quadratic} in the FRG equations~\eqref{eq:flow-eq}, deriving the beta functions for all couplings in it, finding a suitable solution (i.e., an RG trajectory) departing from a UV fixed point and reaching an IR limit compatible with general relativity, and determining the limit $\lim_{k\to 0}g_{C,k}(q^2)$. Every single step in this procedure is technically highly involved. 

\begin{figure}[t] 
	\includegraphics[width=0.5\textwidth]{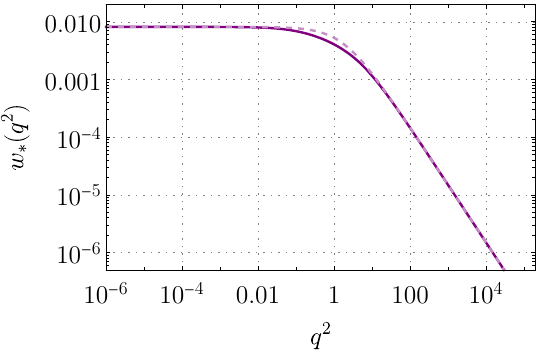} \includegraphics[width=0.49\textwidth]{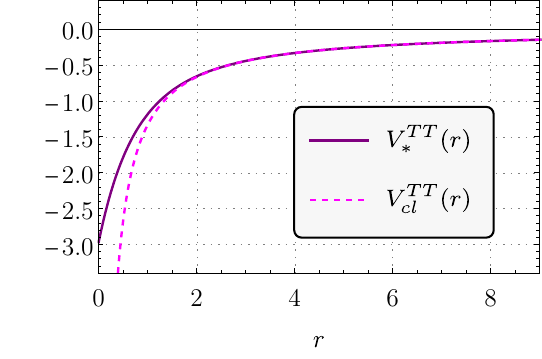}
	\caption{\label{fig:Newt} Dimensionless UV form factor $w_\ast(q^2)$ (left panel) and the corresponding dimensionful TT potential $V_\ast^{TT}(r)$ in coordinate space (right panel). Both are depicted with a solid purple line. The first, is obtained numerically. The analytical approximation of $w_\ast(q^2)$ in Eq.~\eqref{eq:wfit} is also shown for $w_\infty$ in the left panel (dashed, violet line). The second one is obtained by using Eq.~\eqref{eq:potentialfromM} and the TT propagator~\eqref{eq:wfit} with $w_\infty=0.1$. The classical TT potential $V_{cl}^{TT}(r)$ is shown as well for comparison (dashed, magenta line). For $w_\infty>0$, the fixed-point potential $V_\ast^{TT}(r)$ is finite at $r=0$, while the classical one diverges in the same limit, resembling the behavior of $\Phi(r)$ for classical black holes. Both figures have been obtained by using the numerical data in~\cite{Bosma:2019aiu}.}
\end{figure}

In order to make a first evaluation of $V_{qu}(r)$ in asymptotic safety possible,~\cite{Bosma:2019aiu} considered terms in the EAA~\eqref{eq:EAA-quadratic} up to quadratic order in a curvature expansion, neglected the form factor $g_R(\Box)$, exploited an expansion around flat space, and on the right-hand side of the flow equation~\eqref{eq:flow-eq} considered the fluctuations of the conformal mode only. Under these approximations, the projection of the exact FRG equation~\eqref{eq:flow-eq} on the subspace of theories spanned by $\{\Lambda_k,G_k,g_{C,k}(q^2)\}$ yields the beta functions for the corresponding dimensionless couplings $\{\lambda_k,g_k\}$ and dimensionless form factor~$w_k(q^2)=k^{-2}g_{C,k}(q^2)$~\cite{Bosma:2019aiu}. Their non-trivial fixed points are determined by the condition 
\begin{equation}
	k\partial_k g_k=k\partial_k \lambda_k=k\partial_k w_k(q^2)=0\,.
\end{equation}
that is a system of coupled differential and integro-differential equations which can be solved numerically. The Einstein-Hilbert sector spanned by the dimensionless couplings $\{\lambda_k,g_k\}$ turns out to be independent of the last equation and can thereby be solved independently. It admits a non-trivial fixed point at $\{\lambda_\ast= 0.285,g_\ast=0.374\}$ which can act as a UV attractor. Replacing these values in the integro-differential equation for $w_k(q^2)$ one can finally determine the fixed-function~$w_{\ast}(q^2)$ numerically. Fig.~\ref{fig:Newt} shows both the numerical solution (solid, purple line) and its analytic approximation (dashed, pink line),
\begin{equation}\label{eq:wfit}
	w_*^{\mathrm{fit}}\left(q^2\right) \approx w_{\infty}+ \frac{{\kappa}\,\rho}{{\rho}+{\kappa}\,q^2}  \, ,
\end{equation}
where $\rho \approx 0.0149$, $\kappa \approx 0.00817$, and $w_{\infty}$ is a free parameter. Its structure indicates the presence of non-localities in the bare action---a feature that may allow to explain the area law of black holes within a quantum field theoretic setup~\cite{Basile:2021krr}.

Assuming that neither the fluctuations beyond the conformal part nor the flow from the UV fixed point to the physical limit $k\to0$ substantially modify the form of the fixed-point propagator, one may approximate $\mathcal{G}(q^2)$ with the scalar part of the TT graviton propagator at the UV fixed point, and correspondingly
\begin{equation}
	V_{qu}(r)\approx V_{\ast}^{TT}(r)\,.
\end{equation}
This potential can be computed numerically by combining Eq.~\eqref{eq:potentialfromM} with the TT propagator~\eqref{eq:wfit} (cf. Fig.~\ref{fig:Newt}), and is to be confronted with the corresponding TT part of the potential in general relativity,
\begin{equation}
	V_{cl}(r)=-\frac{4}{3}\frac{1}{r}\,.
\end{equation}
At variance of the case of general relativity, and under all approximations detailed above, quantum gravity effects in asymptotic safety make the scattering potential of two scalars finite at $r=0$: if a similar mechanism applies to black hole and cosmological solutions, it would imply that spacetime singularities are weakened in asymptotic safety.

\subsection{Constraints from the gravitational path integral: bare actions and dynamical singularity-resolution mechanism}\label{sect:resolution-path-integral}

In this subsection we summarize the findings in ~\cite{Borissova:2020knn}, where the authors---inspired by the arguments of~\cite{Lehners:2019ibe} in a cosmological context---discuss a ``dynamical'' black hole singularity-resolution mechanism based on the suppression of singular spacetime configurations in the gravitational path integral, and its role in constraining the form of the bare action~$S_{bare}$.

The solutions $g_{\mu\nu}^{sol}$ to the quantum field equations~\eqref{eq:full-eom} are equivalently defined as expectation values according to
\begin{equation}
	g_{\mu\nu}^{sol}\equiv \int \mathcal{D}g\,g_{\mu\nu} e^{i S_{bare}[g]}\,,
\end{equation}
with different solutions corresponding to different initial conditions in the gravitational path integral. Every spacetime configuration $g_{\mu\nu}$ comes with a ``Lorentzian weight'' $e^{i S_{bare}[g]}$. In turn, the latter corresponds to a statistical weight $e^{- S_{bare}[g]}$ in the Euclidean version of the gravitational path integral,
\begin{equation}
	\mathcal{Z}_E\equiv \int \mathcal{D}g^E \,e^{- S_{bare}[g^E]}\,.
\end{equation}
A given solution $g_{\mu\nu}^{sol}$ can be seen as a superposition of spacetime configurations. In particular, the configurations contributing the most are those making the bare action small. By contrast, spacetimes for which the bare action is divergent are suppressed in the Euclidean gravitational path integral; at the Lorentzian level, this suppression translates in a fast-oscillating weighting factor producing destructive interference. This simple consideration can be exploited to investigate the singularity resolution mechanism in quantum gravity~\cite{Lehners:2019ibe,Borissova:2020knn}: in order to produce regular solutions~$g_{\mu\nu}^{sol}$, one needs all singular configurations in the path integral to be suppressed. As a consequence, a theory of quantum gravity can allow for singularity resolution if its bare action diverges when evaluated on singular spacetimes. Based on these arguments, the authors of~\cite{Borissova:2020knn} focused on black hole configurations and investigated what minimal conditions on the bare action $S_{bare}$ are such that this dynamical black hole singularity resolution is possible.

Given a spacetime $g^{sing}_{\mu\nu}$ characterized by a curvature singularity, the divergence of the corresponding bare action $S_{bare}[g^{sing}]$ depends on the curvature invariants appearing in it, as some of them could remain finite despite the singularities of $g^{sing}$. For instance, the Ricci scalar $R$ vanishes for Schwarzschild spacetimes and therefore the Einstein-Hilbert action and its $f(R)$-like extensions do not diverge on a large class of singular spacetimes. This simple consideration implies that higher-derivative terms beyond $f(R)$-like models are required to allow for singularity resolution. In particular, since invariants built on the Ricci tensor $R_{\mu\nu}$ also vanish for Ricci-flat spacetimes, singularity resolution would require at least some terms built from the Riemann tensor, e.g., $R_{\mu\nu\sigma\rho}R^{\mu\nu\sigma\rho}$. A bare action based on Stelle gravity,
\begin{equation}
	S_{bare}[g]=\int d^4 x \sqrt{-g}\,\left(\frac{2\Lambda -R}{16\pi G}+a\,R^2+b\,R_{\mu\nu\sigma\rho}R^{\mu\nu\sigma\rho}\right)\,,
\end{equation}
is therefore the minimal extension beyond Einstein-Hilbert that can in principle allow for singularity resolution~\cite{Borissova:2020knn} (and, in a cosmological context, to allow for a suppression of inhomogeneous and anisotropic configurations, over those satisfying the cosmological principle~\cite{Lehners:2019ibe}) grounded on the dynamical singularity-resolution mechanism described above.

\subsection{Quantum gravity constraints from the principle of least action}\label{sect:least-action-principle-bh}

Lacking a direct derivation from quantum gravity, several spacetime models have been proposed that replace classical singular black holes while reproducing their Schwarzschild exteriors at large distances. These models include, for instance, regular black holes~\cite{Bardeen:1968,Dymnikova:1992ux,Bonanno:2000ep,Hayward:2005gi}, black holes with integrable singularities~\cite{Lukash:2013ts}, wormholes~\cite{Maldacena:2017axo, Marolf:2021kjc, Guo:2021blh}, and a variety of compact objects~\cite{Mazur:2001fv,Mathur:2005zp,Lemos:2007yh, Barcelo:2007yk,Chen:2014loa,Bena:2016ypk}.

As argued in~\cite{Knorr:2022kqp}, a basic requirement for these \emph{ad hoc} models to be physical, i.e., to have a fundamental explanation in terms a complete theory of quantum gravity and matter, is that the corresponding metric $g_{\mu\nu}$ ought to be a solution to effective field equations~\eqref{eq:full-eom} stemming from an effective action~$\Gamma_0$. In particular, one could ask the following question: \emph{given a spacetime described by a metric $\bar{g}_{\mu\nu}$, is there an effective action~$\Gamma_0$ such that}
\begin{equation}
	\left.\frac{\delta \Gamma_0[g]}{\delta g_{\mu\nu}}\right|_{g=\bar{g}} =0\,\,?
\end{equation}
In general, this task is extremely involved. Yet, substantial progress can already be made by focusing on the asymptotic region $r\to\infty$ and by exploiting a relatively general parametrization of the corrections to the Schwarzschild scaling~\cite{Knorr:2022kqp}. This strategy also comes with several advantages:
\begin{itemize}
	\item The constraints resulting from the asymptotic analysis ought to apply to any theory of quantum gravity, since asymptotically all of them have to recover an EFT/QFT description where the principle of least action plays a crucial role.
	\item It allows to constrain a number of different proposed models.
\end{itemize}
In an attempt to answering the aforementioned question, the authors of~\cite{Knorr:2022kqp} derived strong constraints on the asymptotic scaling of modified black holes beyond general relativity by enforcing the validity of the principle of least action. The scope of this subsection is to summarize their work and the resulting constraints.

Following~\cite{Knorr:2022kqp}, we shall focus on the asymptotic scaling of static, spherically-symmetric spacetimes,
\begin{equation}\label{eq:metric}
	\bar{g}_{\mu\nu} = \text{diag} \left(-f_{tt}(r), \frac{1}{f_{rr}(r)}, r^2, r^2 \sin\theta \right) \, ,
\end{equation}
and we shall further assume the metric coefficients to admit an asymptotic expansion of the form
\begin{equation}\label{eq:metricfunctions}
	f_{tt}(r) \sim 1 - \frac{2 G_N m}{r} + \frac{c_t}{r^{n_t}} \, , \qquad f_{rr}(r) \sim 1 - \frac{2 G_N m}{r} + \frac{c_r}{r^{n_r}} \, ,
\end{equation}
where ~$n_r, n_t > 1$, such that $c_ir^{-n_i}$ are sub-leading, asymptotic corrections to the pure Schwarzschild scaling.
This condition applies to the most commonly studied alternatives to Schwarzschild black holes, e.g.,~\cite{Bardeen:1968,Bonanno:2000ep,Hayward:2005gi,Simpson:2018tsi}, as they are build on ratios of polynomials. Specifically, the asymptotic scaling of the Bardeen, Bonanno-Reuter, and Hayward metric coefficients matches the one in Eq.~\eqref{eq:metricfunctions}, with $n_r=n_t=4$ and specific values of $c_r$ and $c_t$. 
By contrast, the asymptotic expansion~\eqref{eq:metricfunctions} does not apply to black holes whose lapse functions contain exponentials of the radial coordinate, e.g., as in~\cite{Dymnikova:1992ux}, or similar non-algebraic functions. The constraints derived in~\cite{Knorr:2022kqp} thus only apply to the former class of models. Note in particular that since the analysis in~\cite{Knorr:2022kqp} is grounded on asymptotic considerations, it applies to any modification to the Schwartschild metric, including singular black holes, and all sorts of black hole mimickers, including  horizonless objects.

At this point, in order to establish whether the metric~\eqref{eq:metric}, for some values of $c_i$ and $n_{i}$, can come from a principle of least action, one can take the following steps:
\begin{enumerate}
	\item Parametrize the effective action $\Gamma_0$, e.g., using a curvature expansion, 
	\begin{equation}\label{eq:EAansatz}
		\Gamma_0 = \frac{1}{16\pi G_N} \int \text{d}^4x \, \sqrt{-g}\, \bigg[ -R - \frac{1}{6} R f_R(\Box) R 
		+R^\mu_{\phantom{\mu}\nu}f_{Ric}(\Box)R^\nu_{\phantom{\nu}\mu} + \mathcal O(\mathcal R^3) \bigg] \, ,
	\end{equation}
	and truncate to a certain order. Here $\Box=-g^{\mu\nu}D_\mu D_\nu$ is the d'Alembert operator built from a metric $g$,~$\mathcal R$ stands for a generic curvature invariant, and $f_i(\Box)$ are form factors, similarly to those we introduced in Eq.~\eqref{eq:eff-action}, that encode the physical momentum dependence of the gravitational couplings~\cite{Christiansen:2014raa, Knorr:2018kog, Bosma:2019aiu, Knorr:2019atm, Knorr:2021niv, Bonanno:2021squ, Fehre:2021eob}. Determining them from first principles ought to be a task of every approach to quantum gravity.
	\item Derive the corresponding field equations~\eqref{eq:full-eom} and evaluate them on the ansatz~\eqref{eq:metric} with metric coefficients~\eqref{eq:metricfunctions}. The resulting equations will dependent on $(c_i,n_i)$ as well as on the couplings $g_i$ in $\Gamma_0$. Structurally, each equation will take the form
	\begin{equation}\label{eq:asy-eqs}
		\sum_{j>\bar{j}} a_j(g_i,c_i,n_i)r^{-j}=0\,,
	\end{equation}
	where $a_j(g_i,c_i,n_i)$ are functions of the couplings $g_i$ and the parameters $(c_i,n_i)$, while $\bar{j}\geq 0$ is an integer whose specific value depends on the form of~$\Gamma_0$.
	\item Establish if there exists $(g_i,c_i,n_i)$ such that the equations are identically fulfilled to leading order in the above asymptotic expansion, $a_{\bar{j}}(g_i,c_i,n_i)=0$. Note that this constitutes only a necessary condition for $\bar{g}_{\mu\nu}$ to be a solution. Next-to-leading order terms ought to be considered in the ansatz~\eqref{eq:metricfunctions} (by construction these terms will not contribute to the leading-order coefficient $a_{\bar{j}}$) and in Eq.~\eqref{eq:asy-eqs}. It turns out that the leading-order condition~$a_{\bar{j}}(g_i,c_i,n_i)=0$ will suffice to put strong constraints on either $(c_i,n_i)$ or $\Gamma_0$.
\end{enumerate}
The first question to ask is whether the modified metric~\eqref{eq:metric} considered in~\cite{Knorr:2022kqp} can be a solution to field equations stemming from a \textit{local} effective action $\Gamma_0$. We shall thus first consider a local version of~\eqref{eq:EAansatz} by expanding the form factors $f_i(\Box)$ in a derivative expansion involving only positive power of the d'Alembertian operator. The resulting effective action thus reads
\begin{equation}\label{eq:EAansatz-Loc}
	\Gamma_0 ^{loc}= \Gamma_{GR}+\Gamma_{\mathcal{R}^2}+\Gamma_{\mathcal{R}^3}+\dots\, ,
\end{equation}
with
\begin{subequations}
	\begin{align}
		& \Gamma_{GR}= -\frac{1}{16\pi G_N} \int \text{d}^4x \, \sqrt{-g}\,R\,, \\
		& \Gamma_{\mathcal{R}^2}= \frac{1}{16\pi G_N} \int \text{d}^4x \, \sqrt{-g}\,\bigg[- \frac{a}{6} R^2 
		+b\,R^\mu_{\phantom{\mu}\nu}R^\nu_{\phantom{\nu}\mu} \bigg] \,, \\
		& \Gamma_{\mathcal R^3} = \frac{1}{16\pi G_N} \int \text{d}^4x \,  \sqrt{-g} \, \bigg[ k_{RC^2} \, R \, C^{\mu\nu\rho\sigma} C_{\mu\nu\rho\sigma} + k_{C^3} \, C_{\mu\nu}^{\phantom{\mu\nu}\rho\sigma} C_{\rho\sigma}^{\phantom{\rho\sigma}\tau\omega} C_{\tau\omega}^{\phantom{\tau\omega}\mu\nu} \bigg] \, .
	\end{align}
\end{subequations}
The condition for the modified metric~\eqref{eq:metric}
to be a solution to the field equations associated with the local effective action~\eqref{eq:EAansatz-Loc} is
\begin{equation}\label{eq:eom-local}
	\left.\frac{\delta \Gamma_0^{loc}[g]}{\delta g_{\mu\nu}}\right|_{g=\bar{g}}=\left.\left(\frac{\delta \Gamma_{GR}[g]}{\delta g_{\mu\nu}}+\frac{\delta \Gamma_{\mathcal{R}^2}[g]}{\delta g_{\mu\nu}}+\frac{\delta \Gamma_{\mathcal{R}^3}[g]}{\delta g_{\mu\nu}} +\dots+\frac{\delta \Gamma_{\mathcal{R}^N}[g]}{\delta g_{\mu\nu}}\right)\,\right|_{g=\bar{g}}=0\, ,
\end{equation}
in an asymptotic expansion for large radii. In particular, the scaling of one of the higher-derivative terms with $r$ ought to cancel general relativity contribution.
Due to the symmetries of~\eqref{eq:metric}, the only two independent components of these field equations can be obtained by deriving every term in $\Gamma_0$ with respect to the metric components $g_{rr}$ and $g_{tt}$. One can thus compute the individual asymptotic contributions of each term in Eq.~\eqref{eq:eom-local}. Setting $n=n_t=n_r$ for shortness, they read
\begin{subequations}\label{eq:eom-contributions}
	\begin{align}
		& \left. \left\{ \frac{\delta\Gamma_{GR}}{\delta g_{tt}}, \frac{\delta\Gamma_{GR}}{\delta g_{rr}} \right\} \right|_{g=\bar{g}} \sim \, \frac{r^{-n-2}}{16\pi G_N} \left\{ c_r(n-1), (c_r-n \, c_t) \right\}\,, \label{eq:eom-gr}\\
		& \left.  \frac{\delta\Gamma^{loc}_{\mathcal{R}^2}}{\delta g_{tt,rr}} \right|_{g=\bar{g}} \sim a_{tt,rr} \, r^{-n-4} \,,\label{eq:eom-r2} \\
		& \left.  \frac{\delta\Gamma^{loc}_{\mathcal{R}^3}}{\delta g_{tt,rr}} \right|_{g=\bar{g}} \sim b_{tt,rr} \, r^{-8} \, ,\label{eq:eom-r3} \\
		& \dots \\
		& \left.  \frac{\delta\Gamma^{loc}_{\mathcal{R}^N}}{\delta g_{tt,rr}} \right|_{g=\bar{g}}  \sim c_{tt,rr} \, r^{-2 - 6 k - 9 l - 2 m} \, ,\label{eq:eom-rn}
	\end{align}
\end{subequations}
where the last term is associated with operators of the form $\left[(\Delta^m R) (C^2)^k (C^3)^l\right]$ or other operators with different contractions and distributions of covariant derivatives. As anticipated, the coefficients $a_{tt,rr}$, $b_{tt,rr}$, and $c_{tt,rr}$, which we called $a_j$ in Eq.~\eqref{eq:asy-eqs}, are functions of~$(n_i,c_i)$ and of the couplings in~$\Gamma_0$. As is evident from these expressions, if one would truncate the effective action $\Gamma_0$ to second order, disregarding terms $\mathcal{O}(\mathcal{R}^3)$, then $\Gamma_0$ would not admit solutions with metric coefficients of the form~\eqref{eq:metricfunctions}. Indeed, there exists no value of $n$ such that the contributions~\eqref{eq:eom-gr} and~\eqref{eq:eom-r2} to the field equations can cancel out. Including terms with six or more derivatives, a cancellation can occur for specific values of $n\geq6$ and $n\neq7$. For instance, including $\mathcal{R}^3$ operators yields the contribution~\eqref{eq:eom-r3} to the field equations, which can cancel the general-relativity one for $n=6$ (and for specific values of the $\mathcal{R}^3$ couplings~\cite{Knorr:2022kqp}). To detail the robustness and generality of these results, it is key to understand the origin of the asymptotic scalings in Eq.~\eqref{eq:eom-contributions}. To this end we notice that:
\begin{itemize}
	\item Due to the nature of the asymptotic expansion of asymptotically-flat spacetimes, all curvature tensors vanish at the expansion point $r\to\infty$. Asymptotically, every curvature tensor thus has to scale as $r^{-h}$ with $h\neq1$. 
	\item The more curvature tensors an operator in $\Gamma_0$ has, the more sub-leading its contribution to the field equations will be.
	\item Since the asymptotic scaling of the Ricci tensor contains $r^{-n}$, operators in $\Gamma_0$ including more than one occurrence of $R_{\mu\nu}$ will not allow to cancel the general-relativity contribution~\eqref{eq:eom-gr} to the field equations for any~$n>0$.
	\item In order to cancel~\eqref{eq:eom-gr} for a given $n$, one needs a leading-order scaling $r^{-h}$ that is independent of $n$. Such a scaling can only come from operators built on Weyl tensors and at most one Ricci tensor or scalar. 
\end{itemize}
Since the first correction to the Schwarzschild scaling for Bardeen, Bonanno-Reuter, and Hayward black holes is $1/r^4$, i.e. $n=4$, these solutions cannot be obtained from a principle of least action based on a local effective action. Enforcing these of other spacetimes with generic $1/r^n$ asymptotic scaling to stem from a principle of least action requires the inclusion of \textit{infrared non-localities}, e.g.,  terms of the form $\mathcal{R}^2 \Delta^{-\frac{n}{2}-2} \mathcal R$, in the effective action $\Gamma_0$~\cite{Knorr:2022kqp}, as these operators contribute to the field equations with terms $\propto r^{-n-2}$ that can cancel the general-relativity scaling~\eqref{eq:eom-gr} for any $n$. Yet, such non-localities would likely lead to observable deviations from GR (see also~\cite{Platania:2023uda}). On the other hand, lapse functions displaying an exponential asymptotic scaling, such as those found in~\cite{Dymnikova:1992ux,Platania:2019kyx,Borissova:2022mgd} and summarized in Sect.~\ref{sect:iterativeRGimpro} and Sect.~\ref{sect:decoupling-mechanism-solutions}, may avoid this conclusion and be compatible with a principle of least action. 

\section{Summary and conclusions}\label{sect:conclu}

The investigation of black holes in asymptotically safe gravity originated in the '90s, with the seminal work by Bonanno and Reuter~\cite{Bonanno:1998ye} on RG-improved black holes. Since then, the big picture of how black holes in an asymptotically safe universe should look like has evolved. In this book chapter we provided a comprehensive chronology of this evolution, from the first works on RG-improved black holes and their generalizations (cf. Sect.~\ref{sect:rg-improved-black-holes}), to recent refinements of the method which attempt to remove its ambiguities (cf. Sect.~\ref{sect:refined-rg-imp}), and more rigorous and solid findings grounded on first-principle calculations in gravitational effective field theory and quantum gravity (cf. Sect.~\ref{sect:bh-from-frg}).

The RG improvement in gravity was introduced as a tool to account for leading-order quantum corrections to classical spacetimes and their dynamics. It consists in promoting the gravitational couplings to RG-scale dependent quantities, and then identifying the artificial RG scale with a physical IR scale. Based on the decoupling mechanism~\cite{Reuter:2003ca}, this procedure ought to provide a short-cut to (an approximation of) the quantum effective action~\cite{Reuter:2003ca} (cf. Sect.~\ref{sect:RG-improv-key-idea}). While some of its most pressing drawbacks---including self-consistency~\cite{Platania:2019kyx}, coordinate dependence~\cite{Held:2021vwd}, and scale setting~\cite{Reuter:2004nv,Reuter:2004nx,Babic:2004ev,Domazet:2012tw,Koch:2014joa,Borissova:2022mgd}---have been partially addressed (cf. Sect.~\ref{sect:refined-rg-imp}), the RG improvement in gravity is to be regarded as a model-building tool to construct quantum-gravity-motivated models. Fully-quantum solutions and their dynamics are to be derived from the quantum effective action.
	
In spite of its limitations, the RG improvement has allowed to build a number of asymptotic-safety-inspired black holes (as well as cosmologies~\cite{Bonanno:2017pkg,Platania:2018eka,Platania:2019qvo}) and to learn general lessons on black holes beyond general relativity (cf.~Sect.~\ref{sect:rg-improved-black-holes}). Within the spherically-symmetric, asymptotically-flat setup, the first implementation of RG improvement yielded the Bonanno-Reuter class of solutions~\cite{Bonanno:1998ye,Bonanno:2000ep,Bonanno:2006eu}. This is a family of regular black holes characterized by two horizons, akin to the well-known Dymnikova~\cite{Dymnikova:1992ux} and Hayward~\cite{Hayward:2005gi} solutions, whose evaporation process is believed to end up in a Planckian remnant. Among the variety of generalizations of Bonanno-Reuter black holes that have been put forth over the years~\cite{Emoto:2005te,Emoto:2006vx,Burschil:2009va,Cai:2010zh,Falls:2010he,Falls:2012nd,Koch:2013rwa,Zhang:2018xzj,Pawlowski:2018swz,Rincon:2020iwy,Chen:2022xjk}, three ingredients are crucial in determining and testing realistic quantum-corrected black holes in asymptotic safety: cosmological constant, spin, and collapse dynamics. Their importance lies in that \emph{(i)} the cosmological constant, in non-unimodular settings, might re-introduce the curvature singularity---unless the dimensionless cosmological constant vanishes at high energies and critical exponents satisfy certain bounds~\cite{Koch:2013owa,Adeifeoba:2018ydh}, \emph{(ii)} non-rotating black holes are unlikely configurations and, moreover, spin is decisive in establishing the shadow properties of black holes beyond general relativity~\cite{Reuter:2006rg,Reuter:2010xb,Held:2019xde}, \emph{(iii)} gravitational collapse renders singularity resolution less straightforward than in the static case, and it is likely that its endpoint be a black hole with an integrable singularity rather than a regular one~\cite{Casadio:2010fw,Fayos:2011zza,Torres:2014gta,Torres:2014pea,Torres:2015aga,Bonanno:2016dyv,Bonanno:2017kta,Bonanno:2017zen,Bonanno:2019ilz}. In particular, this scenario might be desirable as it naturally avoids the potential perturbative instabilities characterizing regular black holes~\cite{Poisson:1989zz,Carballo-Rubio:2018pmi,Bonanno:2020fgp,Carballo-Rubio:2021bpr,Barcelo:2022gii,Carballo-Rubio:2022kad,Bonanno:2022jjp,Carballo-Rubio:2022twq}.

A full understanding of quantum black holes within and beyond asymptotic safety requires going beyond RG improvement and deriving quantum corrections and theoretical constraints from top-down computations. In the last few years much progress has been done in this direction, particularly concerning singularity-resolution mechanisms and general theoretical constraints (cf.~Sect.~\ref{sect:bh-from-frg}).

Singularity resolution within asymptotic safety can be related to the structure of the gravitational path integral~\cite{Lehners:2019ibe,Borissova:2020knn} or, equivalently, to the quantum effective action~\cite{Bosma:2019aiu,Knorr:2022kqp}. As singularities result from the violent collapse of matter under its own gravity, their resolution could rely on an effective weakening of the gravitational interaction at high energies and short distances. Within asymptotic safety, such a weakening is understood in terms of the gravitational anti-screening associated with the Reuter fixed point~\cite{Nink:2012vd}, and may also reflect in scattering amplitudes that are everywhere finite in coordinate space~\cite{Bosma:2019aiu}. At the level of the gravitational path integral, singularity resolution could be related to the suppression of singular configurations via the divergence of the corresponding on-shell action~\cite{Lehners:2019ibe,Borissova:2020knn}. Such a mechanism would require higher derivatives in the bare action.

On top of singularity resolution, providing a fundamental, microscopic explanation of the black hole area law in asymptotic safety is crucial. First steps forward in this direction have been taken in~\cite{Becker:2012js,Becker:2012jx,Koch:2013owa}. Within the Einstein-Hilbert truncation, the scaling of the black hole entropy with the area comes from the structure of on-shell effective action: only boundary terms contribute to the entropy, as the bulk ones vanish on shell. The validity of this mechanism beyond the Einstein-Hilbert truncation is so far unexplored, but if it will turn out to be stable in higher-order truncations---especially those leading to non-Ricci-flat solutions---it could provide a natural explanation for the holographic properties of gravity~\cite{Witten:1998qj,Aharony:1999ti}.

Singularity resolution, stability, and black hole entropy are not the only theoretical requirements constraining physical black holes beyond general relativity. While asymptotic deviations from classical black holes are expected to be tiny, and experimentally untestable, theoretical considerations can put surprisingly strong constraints on them~\cite{Knorr:2022kqp}: requiring the validity of the principle of least action at large distances (i.e., imposing that a given metric is a solution to the dynamics stemming from a gravitational effective action) constrain the asymptotic scaling of black-hole lapse functions beyond Schwarzschild, and even allows to rule out some of the most popular alternatives to Schwarzschild black holes~\cite{Knorr:2022kqp}. This highlights the prominent role of effective actions and the principle of least action in constraining quantum-gravity-inspired models and driving theoretical investigations.

\section*{Acknowledgments}

The author thanks J. Borissova, A. Held, and B. Knorr for comments on various sections of the book chapter, A. Held for many fruitful discussions on the derivations in~\cite{Held:2021vwd}, and B. Knorr for providing the numerical data of~\cite{Bosma:2019aiu} to generate the plots in Fig.~\ref{fig:Newt}.
A.P. acknowledges support by Perimeter Institute for Theoretical Physics. Research at Perimeter Institute is supported in part by the Government of Canada through the Department of Innovation, Science and Economic Development and by the Province of Ontario through the Ministry of Colleges and Universities. A.P. also acknowledges Nordita for support within the ``Nordita Distinguished Visitors'' program and for hospitality during the last stages of development of this work. Nordita is supported in part by NordForsk.

\bibliography{AleBib}

\end{document}